\documentclass[11pt]{article}

\usepackage[utf8]{inputenc}

\usepackage{amsmath}        
\usepackage{amssymb}        
\usepackage{mathtools}      
\usepackage{bm}             
\usepackage{mathrsfs}       
\usepackage{simplewick}     
\usepackage{slashed}        

\usepackage{graphicx}       
\usepackage{epstopdf}       
\usepackage{grffile}        

\usepackage[table]{xcolor}  

\usepackage{array}          
\usepackage{booktabs}       
\usepackage{siunitx}        
\usepackage{multirow}       
\usepackage{dcolumn}        
\newcolumntype{C}[1]{>{\centering\arraybackslash}m{#1}} 
\usepackage{caption}        
\usepackage{subcaption}     
\usepackage{fancyhdr}       
\usepackage{titling}        
\usepackage{rotating}       
\usepackage{lscape}         
\usepackage{fmtcount}       

\usepackage{pgffor}         

\usepackage{cite}           
\usepackage{url}            
\usepackage{hyperref}       

\usepackage[utf8]{inputenc}

\begin{document}

\begin{center}
    \vspace*{15mm}
    \vspace{1cm}
    {\Large \bf Probing the Chern-Simons Portal at the HL-LHC through Displaced Vertices from W Boson Associated Production}

    \vspace{1cm}

    {\bf Mohammad Nourbakhsh$^{1}$ and Mojtaba Mohammadi Najafabadi$^{1,2}$ } \\
    {\small \sl 
        $^{1}$ School of Particles and Accelerators, Institute for Research in Fundamental Sciences (IPM) P.O. Box 19395-5531, Tehran, Iran \\
        $^{2}$ Experimental Physics Department, CERN, 1211 Geneva 23, Switzerland \\
    }
    \vspace*{.2cm}
\end{center}

\vspace*{.2cm}
\vspace*{10mm}

\begin{abstract}\label{abstract}
This study explores the Chern-Simons portal model, an extension of the Standard Model 
that introduces a massive neutral vector boson $X$ associated with a $U_X(1)$ gauge symmetry. 
Motivated by gauge anomaly cancellation, the model incorporates heavy chiral fermions that induce 
observable effects through topological Chern-Simons interactions, despite being inaccessible at 
Large Hadron Collider  energies. We investigate the associated production of the $X$ boson 
with a $W$ boson and jets at the High-Luminosity LHC  with a center-of-mass energy of 14 TeV, 
considering different $X$ masses benchmarks. A multivariate analysis using Boosted Decision Trees 
is employed to separate signal from background processes. 
Detector effects are modeled using a fast simulation tuned to the HL-LHC environment, 
including realistic pile-up conditions with an average of 200 interactions per bunch crossing. 
We derive expected $95\%$ confidence level exclusion limits in two-dimensional parameter 
spaces involving the $X$ boson couplings. 
Our results demonstrate that the HL-LHC can achieve high sensitivity to gauge-anomaly-induced interactions, 
setting robust constraints on the $X$ boson coupling to the $W$ boson down to $\mathcal{O}(10^{-4})$, depending on $m_X$.
\end{abstract}

\newpage

\section{Introduction}\label{sec:introduction}

The Standard Model (SM) of particle physics has proven to be an exceptionally successful 
framework in describing the fundamental constituents of matter and their interactions. 
Its predictions have been thoroughly validated through a wide array of experimental results, culminating in the landmark discovery of the 
Higgs boson at the Large Hadron Collider (LHC)~\cite{ATLAS:2012ae, CMS:2012qbp}. 
Despite its triumphs, the SM is incomplete. It fails to explain several critical phenomena, including the identity and nature of dark matter, 
the origin of neutrino masses, and the observed baryon asymmetry of the universe~\cite{Strassler:2006im, Giudice:1998bp, Dienes:2011ja}. 
These unresolved questions point compellingly to the existence of physics beyond the SM (BSM), motivating extensive theoretical and 
experimental efforts to search for new particles and interactions.

Among the wide range of BSM scenarios, models that predict the existence of long-lived particles (LLPs) 
have emerged as particularly promising avenues for discovery~\cite{Giudice:1998bp, Strassler:2006im, Barbier:2004ez, Alimena:2019zri}. 
LLPs are characterized by decay lengths significantly larger than the typical scale of particle decays ($c\tau \gtrsim 1~\text{cm}$), 
often resulting in displaced decay vertices within the detector. 
These signatures are highly distinctive and benefit from low SM background contamination, 
making them excellent probes at high-energy colliders such as the LHC~\cite{ATLAS:2018tup, CMS:2015pca}. 
Theoretically, LLPs arise naturally in various BSM frameworks including supersymmetric extensions~\cite{Arvanitaki:2012ps, Csaki:2015fea}, 
hidden sector and dark portal models~\cite{Strassler:2006im, Chan:2011aa}, 
and scenarios that address dark matter or neutrino masses through suppressed couplings or mass mixings~\cite{Baumgart:2009tn, Batell:2016zod}. 
The long lifetimes of LLPs typically originate from feeble interactions with the SM, making them challenging to detect but ideally 
suited to the large decay volumes and advanced trigger systems of modern detectors like CMS, ATLAS, and LHCb.

One beyond the SM scenario predicting long-lived particles involves minimal extensions 
of the SM that introduce a new scalar particle ($\Phi$). In such models, $\Phi$ can kinetically mix with the SM Higgs boson, 
leading to production cross sections and decay rates that mirror those of the Higgs boson evaluated at $m_{\Phi}$.
If $m_{\Phi}$ lies below the $B$-meson mass, such particles can be copiously produced in $B$-meson decays~\cite{Ref23}. 
Being neutral under the SM gauge interactions, $\Phi$ may also have a measurable lifetime~\cite{Ref24}, 
giving rise to displaced signatures. The sensitivity of the CERN LHC experiments to these scenarios 
has been investigated in several phenomenological studies~\cite{Ref25,Ref26}, and direct searches 
for long-lived scalars from $B$-meson decays have been carried out by the 
CHARM~\cite{Ref27}, Belle~\cite{Ref28}, BaBar~\cite{Ref29}, and LHCb~\cite{Ref30,Ref31} Collaborations, 
primarily targeting final states where $\Phi$ decays into a pair of charged leptons.
The most recent direct constraints on such LLP scenarios come from dedicated searches at the LHC. 
The CMS Collaboration~\cite{CMS:2025rtd} has searched for LLPs produced in $b$-hadron decays in $pp$ collisions at $\sqrt{s}=13$~TeV, 
using the 2018 dataset corresponding to 41.6~fb$^{-1}$. The analysis targeted LLPs that traverse the detector and interact in the endcap muon system, 
producing hadronic or electromagnetic showers reconstructed as high-multiplicity clusters of hits, in association with a displaced muon. 
No excess above the SM expectation was observed, and stringent limits were set on $\mathcal{B}(B \to K\Phi)$, 
for $m_\Phi$ between $0.3$ and $3.0$~GeV and mean proper decay lengths $c\tau$ in the range 1–500~cm. 
Similarly, the ATLAS Collaboration~\cite{RefATLASLLP} analyzed the full Run~2 dataset of $pp$ collisions at $\sqrt{s}=13$~TeV, 
corresponding to 140~fb$^{-1}$, to search for LLPs with masses 5–55~GeV decaying hadronically in the inner detector. 
Benchmark scenarios included exotic Higgs boson decays to LLP pairs and long-lived axion-like particles (ALPs) produced in association with a vector boson. 
No significant excess was observed, leading to upper limits on the Higgs boson branching fraction to LLP pairs, the ALP production cross section, 
and, for the first time, the branching fraction of the top quark to an ALP and a $u$- or $c$-quark.

Searches for LLPs have also been investigated at future lepton colliders such as FCC-ee \cite{rf1,rf2}. 
Studies based on $e^+e^- \to ZH$ production show that displaced-vertex searches for exotic 
Higgs decays into long-lived dark scalars could probe branching ratios down to $10^{-4}$, 
with machine learning techniques pushing the sensitivity to $9.7\times 10^{-7}$ while 
maintaining high background rejection.

These experimental results, together with the growing theoretical interest in gauge-anomaly-induced portals, 
motivate a dedicated study of novel LLP production modes and signatures, as pursued in this work.
A particularly intriguing framework that gives rise to LLP signatures is the Chern-Simons portal model~\cite{Antoniadis:2009ze}. 
This scenario extends the SM by introducing a new massive neutral vector boson, denoted $X$, associated with an additional $U_X(1)$ gauge symmetry. 
The model is constructed to maintain gauge invariance in the presence of new chiral fermions that induce mixed 
gauge anomalies; specifically, anomalies such as $U_X(1)\text{-}SU(2)_L^2$ that must be canceled for theoretical consistency. 
In this context, the cancellation of anomalies is achieved not by the fermion content alone but through effective 
Chern-Simons interactions that emerge after integrating out heavy fermions. 
Unlike most heavy new physics scenarios that decouple at low energies~\cite{Appelquist:1974tg}, 
these Chern-Simons terms persist due to their topological origin, leading to unsuppressed and potentially observable effects even if the underlying fermions are inaccessible at LHC energies. 
This non-decoupling behavior renders the $X$ boson phenomenologically accessible at the LHC, especially in searches for long-lived particles.

In the Chern-Simons portal model, the $X$ boson may possess a macroscopic lifetime, resulting in displaced decays into SM particles such as muon pairs. 
These decays proceed through anomaly-induced interactions parameterized by the coupling $c_W$ in the $XWW$ vertex and by a 
small axial-vector coupling $g_{Xll}$ to leptons~\cite{Antoniadis:2009ze, Bondarenko:2019tss}. 
The High-Luminosity LHC (HL-LHC), with a projected integrated luminosity of 3000 fb$^{-1}$ and a center-of-mass energy of 14 TeV, 
provides an ideal environment to probe such displaced signatures, particularly through the enhanced capabilities of the CMS muon system~\cite{CMS:2015pca}. 
Recent advances in displaced muon triggering significantly extend the reach for LLP searches. 
For instance, the CMS experiment has developed upgraded triggers; such as the Kalman Beam Muon Track Finder (kBMTF); 
that are specifically designed to identify displaced muons with large impact parameters ($d_0 \gtrsim 10~\text{cm}$). 
These triggers exploit sub-nanosecond timing resolutions (approximately 2–3 ns) to suppress prompt background contamination 
and enable efficient selection of non-standard decay topologies. 
These capabilities make displaced dimuon searches among the most sensitive channels for probing new vector bosons in models like the Chern-Simons portal.

In this work, we investigate the HL-LHC's sensitivity to the Chern-Simons portal model via the associated production of the $X$ boson 
with a $W$ boson and at least one hard jet in proton-proton collisions. 
We focus on the decay chain $p p \to W^\pm + j + X$, followed by $X \to \mu^+\mu^-$ and $W^\pm \to e^\pm \nu_e$, yielding a 
final state featuring a prompt electron, missing transverse energy, a displaced dimuon pair, and at least one hadronic jet.  
This channel is especially promising due to the large gluon parton distribution functions (PDFs) at LHC energies, 
which enhance the $gq \to W^* q \to WXq$ contribution, and because the presence of a jet facilitates triggering and background suppression. 
A multivariate analysis employing Boosted Decision Trees (BDTs) is performed to distinguish signal from Standard Model backgrounds, 
and $95\%$ confidence level exclusion limits are derived for benchmark $X$ boson masses of 5, 10, and 15 GeV.  
The analysis incorporates full matrix-element-level simulations and a realistic treatment of detector effects, 
together with systematic uncertainties from parton distribution functions (PDFs), renormalization and factorization scales, 
the strong coupling constant, and integrated luminosity, while for other sources that cannot be treated explicitly, overall values have been taken into account. 
Our results underscore the potential of displaced vertex searches at the HL-LHC to probe anomaly-driven interactions and constrain novel extensions of the SM.

This paper is organized as follows.  
Section~\ref{sec:theoretical} provides a brief overview of the theoretical framework of the Chern--Simons portal model.  
The analysis strategy, including event generation and simulation, background estimation, and the discrimination of 
signal from background using a multivariate approach, is presented in Section~\ref{sec:analysis}.  
The statistical method and results are discussed in Section~\ref{sec:results}.  
Finally, Section~\ref{sec:conclusion} summarizes the findings and presents the conclusions.

\section{Theoretical Framework}\label{sec:theoretical}

The Chern-Simons portal model extends the SM by introducing a new massive neutral vector boson, 
denoted $X$, associated with an additional abelian gauge symmetry $U_X(1)$. 
This model is motivated by the desire to probe new physics beyond the SM, particularly through the framework of gauge anomalies and their possible resolution mechanisms. 
Gauge anomalies are quantum inconsistencies that arise when certain triangle 
Feynman diagrams involving chiral fermions do not preserve gauge invariance, leading to violations of current conservation at the quantum level. 
In a consistent theory, such anomalies must cancel. 
The SM achieves this cancellation via the particular charge assignments of its fermion content. 
However, in many extensions of the SM, new fermions are introduced whose anomaly contributions must be canceled by additional mechanisms.
The Chern-Simons portal model considers this scenario and introduces heavy fermions that are chiral under 
either the SM gauge group $SU(3)_C \times SU(2)_L \times U(1)_Y$, or the new $U_X(1)$. 
These heavy fermions induce mixed gauge anomalies such as $U_X(1)SU(2)^2$ 
(i.e., anomalies involving the $X$ gauge boson and two $W$ bosons), which, although canceled at the full theory level, 
leave behind effective interactions in the low-energy theory once the heavy fermions are integrated out. 
These residual effects take the form of Chern-Simons terms, which are topological in nature and thus remain unsuppressed at low energies, 
in contrast to typical decoupling behavior. This feature makes them particularly appealing for collider searches, 
as their phenomenological consequences can be significant even if the underlying new physics scale is high.

\begin{figure}[ht]
	\centering
	\includegraphics[width=0.4\linewidth]{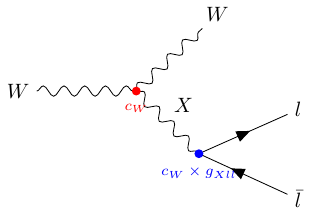}
	\caption{Representative Feynman diagram displaying the $WWX$ and $X\ell\ell$ vertices, corresponding to the couplings $c_W$ and $g_{Xll}$.}
	\label{fig:wwx}
\end{figure}

The low-energy effective Lagrangian arising from these anomaly-induced terms is given by~\cite{Antoniadis:2009ze}:
\begin{equation}
    \mathcal{L}_{CS} = c_W\epsilon^{\mu\nu\lambda\rho} X_\mu W_\nu \partial_\lambda W_\rho
    + c_{\gamma} \cos\theta_W \epsilon^{\mu\nu\lambda\rho} X_\mu Z_\nu \partial_\lambda A_\rho +
    c_{Z} \sin\theta_W \epsilon^{\mu\nu\lambda\rho} X_\mu Z_\nu \partial_\lambda Z_\rho,
    \label{eq:chern-simons}
\end{equation}
where the fields $A_\mu$, $Z_\mu$, and $W_\mu$ represent the photon, $Z$ boson, and $W^\pm$ bosons, 
respectively, and $\epsilon^{\mu\nu\lambda\rho}$ is the fully antisymmetric Levi-Civita tensor. 
The coefficients $c_W$, $c_\gamma$, and $c_Z$ are dimensionless Wilson coefficients encoding the strength of 
the anomaly-induced interactions, and are model-dependent quantities determined by the charge assignments and multiplicities of the heavy fermions.
To allow the $X$ boson to decay within the detector and produce visible signatures, it must also couple to SM matter. In this work, we assume it interacts primarily with charged leptons. 
The interaction of $X$ with leptons is parameterized by the effective axial-vector Lagrangian:
\begin{equation}
    \mathcal{L}_{Xll} = c_{W}g_{Xll} X^{\nu}\sum_{l = e,\mu,\tau}\bar{l}\gamma_{5}\gamma_{\nu}l,
    \label{eq:interaction-with-leptons}
\end{equation}
where $g_{Xll}$ is a dimensionless coupling constant. This structure ensures that the interaction is parity-violating and CP-odd, consistent with the nature of the Chern-Simons origin. 
A representative Feynman diagram showing the vertices corresponding to the $XWW$ and $X\ell\ell$ interactions is displayed in Fig.~\ref{fig:wwx}.

The model parameters $c_Z$, $c_W$, and $c_\gamma$ are constrained by electroweak precision observables and by the total decay widths of the SM gauge bosons. 
In particular, new decay channels such as $Z \to \gamma X$ and $W \to X f\bar{f}'$ can be induced via the effective Lagrangian~\eqref{eq:chern-simons}, leading to additional contributions to the widths of the $Z$ and $W$ bosons. 
The dominant contributions in the small $m_X$ regime come from the longitudinal component of the $X$ boson, due to the enhancement factor $M_{V}^2 / m_X^2$ (with $V = Z, W$), which compensates for the small coupling.

These contributions have been computed in~\cite{Alekhin:2015byh}, and are given approximately by:
\begin{equation}
    \Gamma(Z\rightarrow\gamma X) = \frac{c_{\gamma}^2 \cos\theta_W M_W}{96\pi} 
    \left(\frac{M_Z^2}{m_X^2} + 1\right), \quad
    \Gamma(W^+\rightarrow Xu\bar{d}) \approx
    \frac{c_W^2 \alpha_W M_W}{432 \pi^2}
    \frac{M_W^2}{m_X^2},
\end{equation}
where $\alpha_W$ is the electroweak coupling constant and $\theta_W$ is the Weinberg angle. A similar expression holds for the decay $Z \rightarrow Z^* X$.
From these expressions, one sees that the decay widths scale as $\sim c_V^2 / m_X^2$ in the limit $m_X \ll M_V$. Therefore, to remain consistent with the measured values of the $Z$ and $W$ widths, the couplings must obey the following bounds:
\begin{equation}
    c_Z^2,\: c_W^2 \lesssim 10^{-3}\left(\frac{m_X}{\text{GeV}}\right)^2, \quad
    c_\gamma^2 \lesssim 10^{-9}\left(\frac{m_X}{\text{GeV}}\right)^2,
\end{equation}
The significantly tighter constraint on $c_\gamma$ arises from LEP bounds on single-photon events~\cite{L3:1997exg}, which strongly limit the exotic decay $Z \to \gamma + \text{invisible}$.

In our analysis, following~\cite{Bondarenko:2019tss}, we consider the benchmark regime where $c_\gamma, c_Z \ll c_W$, 
such that the phenomenology is dominated by the $XWW$ interaction. 
This assumption is both theoretically well-motivated and experimentally justified, given the comparative 
looseness of the bounds on $c_W$ and the accessibility of the $W + X$ production channel at hadron colliders. 
Once produced, the $X$ boson is assumed to decay visibly via its interaction with muons: $ X \to \mu^+ \mu^-,$
as described by the Lagrangian in Eq.~\eqref{eq:interaction-with-leptons}. However, due to its coupling 
to $W$ bosons, the $X$ boson can also decay via purely gauge-mediated channels to four fermions.
The ratio of the four-body to two-body decay widths has the following form \cite{Bondarenko:2019tss}:
\begin{equation}
\frac{\Gamma_{X \to \text{four-femions}}}{\Gamma_{X \to \mu\mu}}  \simeq \frac{10^{-7}}{g_{X\mu\mu}^{2}}\left(\frac{m_{X}}{40\text{ GeV}}\right)^{8}.
\end{equation}
This ratio shows that for $m_X \lesssim 40$ GeV and $g_{X\mu\mu}^2 \gtrsim 10^{-7}$, the gauge-mediated four-fermion decay is highly 
suppressed relative to the two-body decay into muons. As a result, the dominant decay channel of the $X$ boson in this mass and coupling regime is into 
a dimuon pair, justifying the displaced dimuon final state as the optimal discovery signature.

\section{Analysis}\label{sec:analysis}

In this section, we outline the analysis methodology. We search for the signal process $p + p \to W^\pm + j + X$ (plus the charge-conjugated final states), 
with the subsequent decays $W^\pm \to e^\pm \nu_e$  and $X \to \mu^+ \mu^-$. 
In this analysis we focus on the $W$ boson decaying to electrons and the muonic $W$ decay mode is not included. This choice avoids overlap between prompt muons from 
the $W$ decay and the displaced muons from $X$, thereby reducing the combinatorial background that 
would arise in final states with three muons. While in the real experimental conditions, the analysis 
relies on the very efficient muon triggering expected 
from the CMS Phase-2 upgrades, the use of the electron channel is further motivated by 
the excellent electron triggering and reconstruction capabilities provided by the High Granularity 
Calorimeter (HGCAL) and the upgraded tracker~\cite{CMS:2017gat}.

\begin{figure}[ht]
    \centering
    \includegraphics[width=0.3\textwidth]{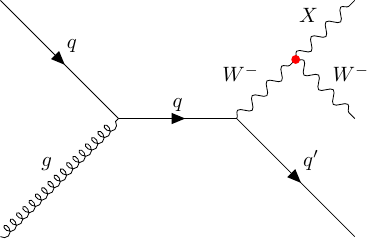}
    \includegraphics[width=0.3\textwidth]{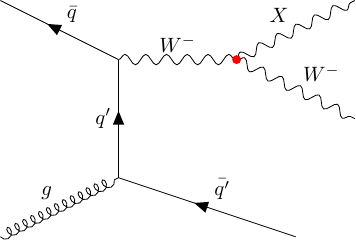}
    \includegraphics[width=0.3\textwidth]{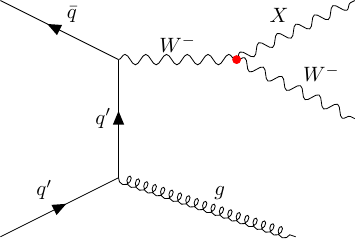}\\
    \caption{\small Representative Feynman diagrams contributing to the leading-order associated production of the $X$ gauge boson with a jet and a $W$ boson in proton--proton collisions.}
    \label{fig:signal-feynman-diagrams}
\end{figure}

The Feynman diagrams corresponding to the signal process are shown in Fig.~\ref{fig:signal-feynman-diagrams}.
BSM interactions are represented by red filled circles in the Feynman diagrams. These processes are described by the effective Lagrangian in Eq.~\ref{eq:chern-simons}.

In an earlier study such as~\cite{Bondarenko:2019tss}, the production of the $X$ boson was considered through the process $q + \bar{q}' \rightarrow W^* \rightarrow W + X$. 
While this channel is theoretically clean and straightforward, it relies on the $s$-channel annihilation of valence or sea quarks, which is suppressed at the LHC due to 
the relatively small quark-antiquark luminosity, particularly for non-valence flavors like $\bar{q}'$. This significantly limits the production rate of $X$ bosons in this channel, 
especially at high luminosity where gluon-initiated processes dominate.
In our analysis, we consider a broader and more inclusive set of partonic subprocesses that probe the anomaly-induced $W\text{-}W\text{-}X$ interaction 
via the decay $W^* \rightarrow W + X$. These include both $s$-channel and $t$-channel diagrams initiated by different combinations of quarks and gluons, 
which are expected to dominate at LHC energies due to their enhanced parton luminosities.
A dominant contribution arises from a gluon--quark initial state:
$g + q \rightarrow W^* + q \rightarrow W + X + q,$
where both $s$- and $t$-channel diagrams contribute. This production mechanism benefits from the large gluon PDF at the LHC and results in a final-state quark (jet) recoiling against the $W + X$ system. This topology not only enhances the signal production rate but also provides experimental advantages, such as improved trigger efficiency and better separation from backgrounds involving prompt leptons and jets.
Another relevant class of diagrams involves quark--antiquark annihilation into a gluon and an off-shell $W$ boson:
$q + \bar{q}' \rightarrow g + W^* \rightarrow g + W + X,$
which also probes the Chern-Simons vertex. Although the $q\bar{q}'$ luminosity is smaller than that of the $gq$ initial state, 
this contribution is non-negligible at higher partonic center-of-mass energies and contributes to the overall sensitivity. 
Both production topologies are incorporated into our matrix-element-level event generation, ensuring a realistic and 
comprehensive modeling of the signal. By including these multiple production channels, our study provides a more complete and robust assessment of the HL-LHC's sensitivity to the Chern-Simons portal model.

\subsection{Signal and background simulation}
\label{subsec:selection}

The analysis begins with the generation of signal and background events. In these simulated events, physics objects such as jets, electrons, and muons are reconstructed, and selection criteria are applied to isolate signal-like events. The selection strategy and discriminating variables are optimized to maximize sensitivity while maintaining control over Standard Model background contributions.
Assuming the HL-LHC operates at a center-of-mass energy of 14 TeV, we investigate this process and derive upper bounds on the parameters $c_W$ and $g_{X\ell\ell}$ for $X$ boson masses of 5, 10, and 15 GeV. The benchmark couplings used in the simulation are $c_W^2 = 5 \times 10^{-8}$ for all three mass points and $g_{X\ell\ell}^2 = 10^{-5}$, $10^{-6}$, and $10^{-7}$, respectively.
The effective Lagrangian in Eq.~\ref{eq:chern-simons} and Eq.~\ref{eq:interaction-with-leptons} was implemented in \textsc{FeynRules}~\cite{Alloul:2013bka}, and the corresponding Universal FeynRules Output (UFO)~\cite{Degrande:2011ua} was used in \textsc{MadGraph5\_aMC@NLO}~\cite{Alwall:2011uj} to compute partonic-level cross sections and generate hard-scattering events for both signal and background processes. The NNPDF23~\cite{Ball:2012cx} parton distribution functions were used with the center-of-mass energy set to 14 TeV. \textsc{Pythia 8.2.43}~\cite{Sjostrand:2014zea} was used to handle parton showering, hadronization, and the decays of unstable particles.

Showered events were processed using \texttt{Delphes 3.4.2}~\cite{deFavereau:2013fsa} for fast detector simulation, using the official CMS Phase-II detector card \texttt{CMS\_PhaseII\_200PU.tcl}\footnote{\url{https://github.com/delphes/delphes/blob/master/cards/CMS_PhaseII/CMS_PhaseII_200PU.tcl}} provided in the Delphes distribution. This configuration models the High-Luminosity LHC (HL-LHC) running conditions, where the instantaneous luminosity is projected to reach $7.5 \times 10^{34}~\mathrm{cm^{-2}s^{-1}}$. Under these conditions, each bunch crossing will experience on average 200 simultaneous proton-proton interactions, known as "pile-up" (PU). 
This high-pile-up environment presents major challenges for physics object reconstruction, particularly in identifying isolated leptons, reconstructing jets, and measuring missing transverse energy. The Phase-II Delphes card includes upgraded tracker granularity, calorimeter segmentation, and realistic pile-up mitigation algorithms that reflect the design and performance goals of the CMS detector upgrade. 

In our simulation, pile-up is accounted for in both jet reconstruction and lepton isolation using Delphes’s pile-up subtraction tools, ensuring that signal efficiencies and background contaminations reflect HL-LHC conditions. This realistic treatment is crucial, especially given the role of soft and collimated dimuon pairs from long-lived $X$ boson decays, which could be sensitive to detector granularity and PU effects. 
Signal samples were generated at leading order using the UFO model\footnote{\url{https://feynrules.irmp.ucl.ac.be/attachment/wiki/ChernSimonsPortal/ChernSimons_UFO.zip}}. 
All SM background processes were also generated at leading order using consistent parton-level cuts and PDF sets.

Jets are reconstructed using the anti-$k_t$ algorithm~\cite{Cacciari:2008gp} as implemented in \texttt{FastJet~3.3.2}~\cite{Cacciari:2011ma}, 
with a radius parameter $R = 0.4$ defining the clustering distance in the rapidity–azimuthal angle ($y$–$\phi$) plane, where $\Delta R = \sqrt{(\Delta y)^2 + (\Delta \phi)^2}$. 
Reconstructed jets are required to have transverse momentum $p_T > 30$~GeV and pseudorapidity $|\eta| < 5.0$.  
Isolated electrons are selected using a pileup-corrected relative isolation variable, $I_{\mathrm{rel}}^{\mathrm{corr}}$, defined as:
\begin{equation}
    I_{\mathrm{rel}}^{\mathrm{corr}} = \frac{\sum p_T^i - \rho \cdot A_{\mathrm{eff}}}{p_T^{\mathrm{P}}},
\end{equation}
where $p_T^{\mathrm{P}}$ is the transverse momentum of the electron candidate, and the sum $\sum p_T^i$ runs 
over the transverse momenta of all reconstructed particles (excluding the candidate) with $p_T > 0.5$~GeV within 
a cone of radius $\Delta R = 0.3$ in the ($\eta$–$\phi$) plane. The pileup correction term $\rho \cdot A_{\mathrm{eff}}$ 
accounts for the average event energy density $\rho$, scaled by the effective isolation area $A_{\mathrm{eff}}$. 
An electron is considered isolated if $I_{\mathrm{rel}}^{\mathrm{corr}} < 0.15$. Selected electrons are further required to have $p_T > 30$~GeV and $|\eta| < 3.0$.

Muon isolation is treated differently due to the expected signal topology. In the process $X \to \mu^+\mu^-$, the muon pair is often 
soft and highly collimated, similar to dimuons from $J/\psi$ decays. Following Ref.~\cite{CMS:2022fsq}, for muons 
separated by $\Delta R < 0.3$, the scalar sum of the transverse momenta of surrounding particles (excluding the muons) 
must be less than 50\% of the leading muon’s $p_T$. For muons outside this cone, each must independently satisfy the same 
isolation requirement. Selected muons must have $p_T > 3$~GeV and $|\eta| < 2.8$.  
Events are required to contain exactly two isolated, oppositely charged muons, exactly one isolated electron (or positron), 
and at least one jet passing the above requirements. The missing transverse momentum, $\slashed{E}_T$, 
defined as the magnitude of the vector sum of all reconstructed particle momenta in the transverse plane, 
must exceed 30~GeV. Jets within $\Delta R < 0.4$ of any selected lepton are removed to prevent double-counting. 
Fake leptons and jet lepton misidentification are not simulated in our fast detector framework (\texttt{Delphes}) and are therefore excluded from the present study.

\subsection{Background processes}

Since the signal signature consists of exactly two oppositely charged muons, one isolated electron (or positron), 
and large missing transverse momentum ($\slashed{E}_T$), several SM processes can yield 
indistinguishable visible final states, thereby constituting irreducible backgrounds. 
The dominant contributions originate from diboson ($VV$) and triboson ($VVV$) production, where $V = W^{\pm}, Z$. 
In $WZ$ and $ZZ$ production, prompt muons arise from the $Z \to \mu^+ \mu^-$ decay, while the electron 
and $\slashed{E}_T$ are provided by the leptonic decay of the $W$ boson. Triboson channels such as $WWZ$, $WZZ$, or $WWW$ 
can produce the required lepton multiplicity through combinations of vector boson decays. Single-top production in the 
$tW$ channel and associated $tZj$ production can mimic the signal topology when the $Z$ boson decays into a 
dimuon pair and the $W$ boson decays to an electron and a neutrino, with jets from the top-quark decay present in the event. 
Similarly, $t\bar{t}W^{\pm}$ events can yield two muons and one electron for instance when one $W$ boson from the top-pair decays 
leptonically to a muon, the other top decay produces another muon via $W \to \mu \nu$, and the associated $W$ boson 
decays to an electron. Associated Higgs-vector boson production ($hV$) also contributes also as background, leaving the signal-like final state. 
Finally, $W^{\pm} \mu^{+} \mu^{-}$ production directly yields the same lepton flavor and multiplicity as the signal when the $W$ boson 
decays to an electron and neutrino, and the dimuon system originates from a virtual photon.

A particularly challenging background arises from processes in which a $W$ boson is produced in association with 
heavy-flavor quarks ($b$ or $c$), followed by heavy-flavor hadron decays that yield a displaced dimuon pair. 
In such events, the $W$ boson decays leptonically, providing the prompt isolated electron and large $\slashed{E}_T$ from 
the neutrino, while the heavy-flavor quarks hadronize into $B$ or $D$ mesons whose decay products include two oppositely charged muons. 
When the muons are produced with small angular separation and originate from a displaced vertex, the topology closely mimics 
the signal decay $X \to \mu^+\mu^-$ both in spatial geometry and in kinematic distributions. Two main production mechanisms 
for the dimuon system are relevant: \textit{(i) Direct semileptonic channels:} both muons are produced directly in the decay 
chains of $B$ or $D$ mesons, without an intermediate quarkonium state. This occurs, for example, when both heavy-flavor 
hadrons undergo semileptonic decays ($H_Q \to \mu + \nu_\mu + X_{\mathrm{had}}$), producing two muons that can occasionally be 
collimated due to the boost of the parent hadrons. These contributions are modeled by
 generating $pp \to W + b\bar{b}$ and $pp \to W + c\bar{c}$ at the parton level, followed by showering, hadronization and full decay simulation. 
 \textit{(ii) Quarkonium-mediated channels:} the dimuon system originates from the decay of a $J/\psi$ meson ($c\bar{c}$ bound state) 
 produced in heavy-flavor decays or directly from $c\bar{c}$ hadronization:
\begin{itemize}
    \item Non-prompt $J/\psi$ from $B$ decays: $B \to J/\psi + X_{\mathrm{had}}$, followed by $J/\psi \to \mu^+ \mu^-$. This is the 
    dominant source in $W + b\bar{b}$ events, where the $J/\psi$ inherits a sizeable displacement from the long $B$-hadron lifetime.
    \item Prompt $J/\psi$ from $c\bar{c}$ production: In $W + c\bar{c}$ events, the $c$ and $\bar{c}$ quarks can hadronize directly into 
    a $J/\psi$ meson or through feed-down from higher charmonium states. The $J/\psi$ then decays promptly to $\mu^+\mu^-$, with possible 
    displacement arising only if the muons originate from secondary charm decays instead.
\end{itemize}
The specific channels considered in the simulation are:
\begin{align*}
    & pp \to W^{\pm} b \bar{b},\quad W^{\pm} \to e^{\pm} \nu_e,\quad B \to J/\psi + X_{\mathrm{had}},\quad J/\psi \to \mu^+ \mu^-, \\
    & pp \to W^{\pm} c \bar{c},\quad W^{\pm} \to e^{\pm} \nu_e,\quad c\bar{c} \to J/\psi + X_{\mathrm{had}},\quad J/\psi \to \mu^+ \mu^-.
\end{align*}

The generation strategy explicitly separates \textit{direct} and \textit{quarkonium-mediated} dimuon production to 
ensure accurate modeling of the muon pair kinematics and displacement distributions. Parton-level processes
 are generated with \texttt{MadGraph5\_aMC@NLO}, while heavy-flavor hadronization, meson decays, 
 and $J/\psi \to \mu^+\mu^-$ branching are simulated in \texttt{Pythia~8}. This setup captures the full angular 
 correlations, boosts, and lifetimes of the parent hadrons, allowing us to correctly reproduce the 
 small-$\Delta R$, displaced dimuon signatures that can fake the signal. Such dedicated treatment is 
 essential: generic inclusive $W$ + jets samples tend to underestimate the probability of producing
  events where both muons are simultaneously displaced, collimated, and accompanied by a prompt 
  isolated electron with large $\slashed{E}_T$. By isolating and simulating these heavy-flavor channels explicitly, 
  the analysis achieves a reliable background estimate for one of the most signal-like topologies in the search. 
  Applying the baseline selection to both the signal and background samples yields the efficiencies shown in Table~\ref{tab:eff_bkg} and Table~\ref{tab:eff_sig}.

\begin{table}[t]
    \centering
    \begin{tabular}{l c}
        \toprule
        Process & Efficiency [\%] \\
        \midrule
        $hW$              & 1.60 \\
        $t\bar{t}W$       & 1.47 \\
        $t\bar{t}$        & 1.09 \\
        $tW$              & 1.08 \\
        $WWW$             & 0.96 \\
        $W J/\psi$        & 0.52 \\
        $WW$              & 0.27 \\
        $tZj$             & 0.21 \\
        $WWZ$             & 0.12 \\
        $W B/D$           & 0.10 \\
        $hZ$              & 0.09 \\
        $W \mu^+ \mu^-$   & 0.04 \\
        $ZZ$              & 0.02 \\
        \bottomrule
    \end{tabular}
    \caption{Selection efficiencies for SM background processes after all baseline cuts.}
    \label{tab:eff_bkg}
\end{table}

\begin{table}[t]
    \centering
    \begin{tabular}{l c}
        \toprule
        Signal hypothesis & Efficiency [\%] \\
        \midrule
        $m_X = 5~\mathrm{GeV}$  & 21.64 \\
        $m_X = 10~\mathrm{GeV}$ & 24.02 \\
        $m_X = 15~\mathrm{GeV}$ & 23.01 \\
        \bottomrule
    \end{tabular}
    \caption{Selection efficiencies for the signal processes after all baseline cuts. 
    Signal samples assume $c_W^2 = 5 \times 10^{-8}$ and $g_{X\ell\ell}^2 = 10^{-5}$, $10^{-6}$, and $10^{-7}$ for $m_X = 5$, $10$, and $15~\mathrm{GeV}$, respectively.}
    \label{tab:eff_sig}
\end{table}

Following Ref.~\cite{Bondarenko:2019tss}, we apply an additional displaced-vertex (DV) selection to enhance 
sensitivity to the LLP signal. The CMS muon tracker geometry restricts reconstructible decays to transverse and 
longitudinal displacements $l_{\mathrm{max},\perp} < 3$~m and $l_{\mathrm{max},\parallel} < 7$~m, respectively. 
To suppress residual backgrounds, especially from heavy-flavor decays, we require $l_{\mathrm{DV}} > 2$~cm from the primary vertex. 
This requirement effectively removes most prompt decays and a large fraction of $b$- and $c$-hadron decays, while retaining good acceptance 
for long-lived $X$ bosons.

\begin{figure}[!t]
    \centering
    \begin{subfigure}[b]{0.48\textwidth} 
    \centering
    \includegraphics[width=\textwidth]{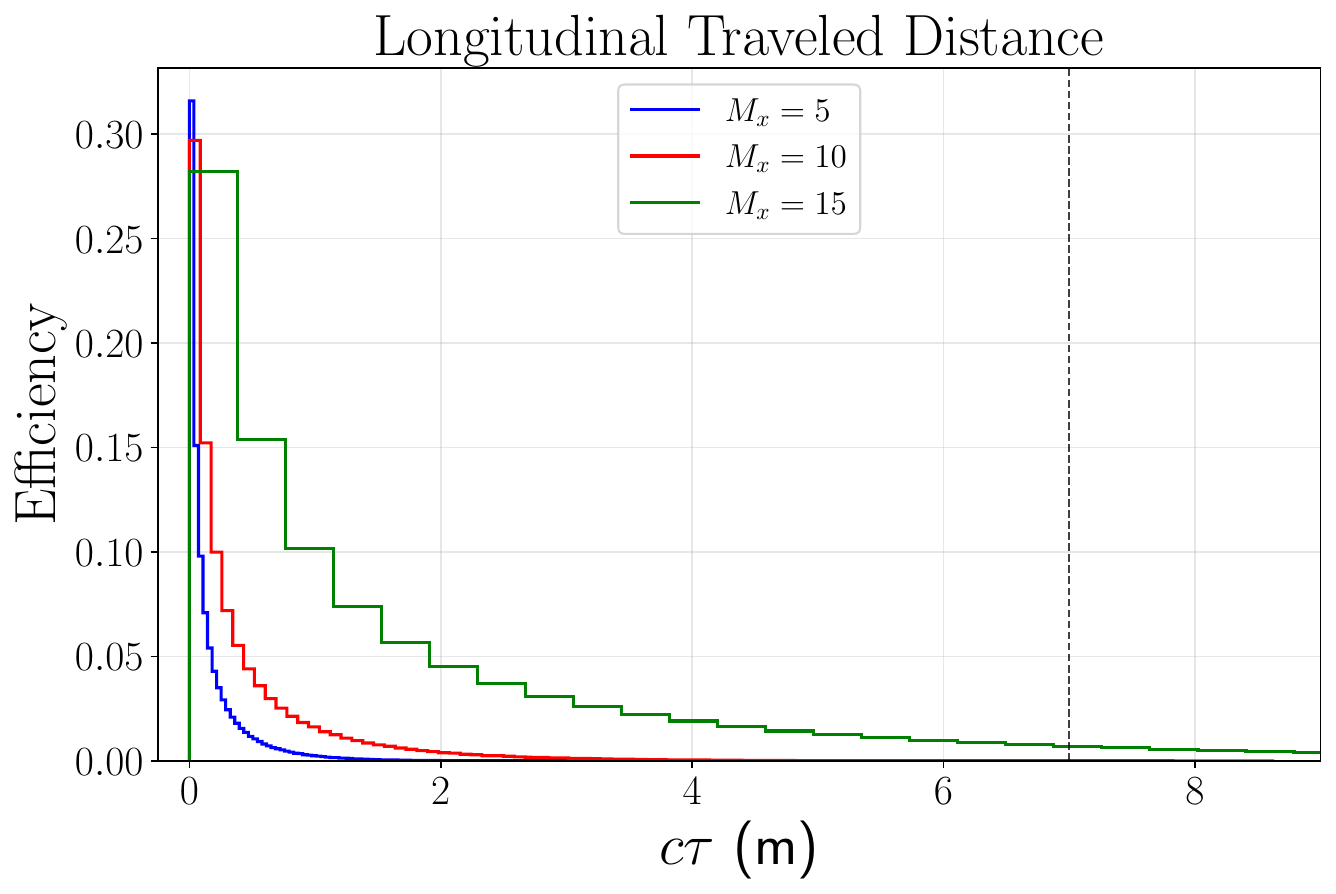}
    \end{subfigure} 
    \begin{subfigure}[b]{0.48\textwidth} 
    \centering
    \includegraphics[width=\textwidth]{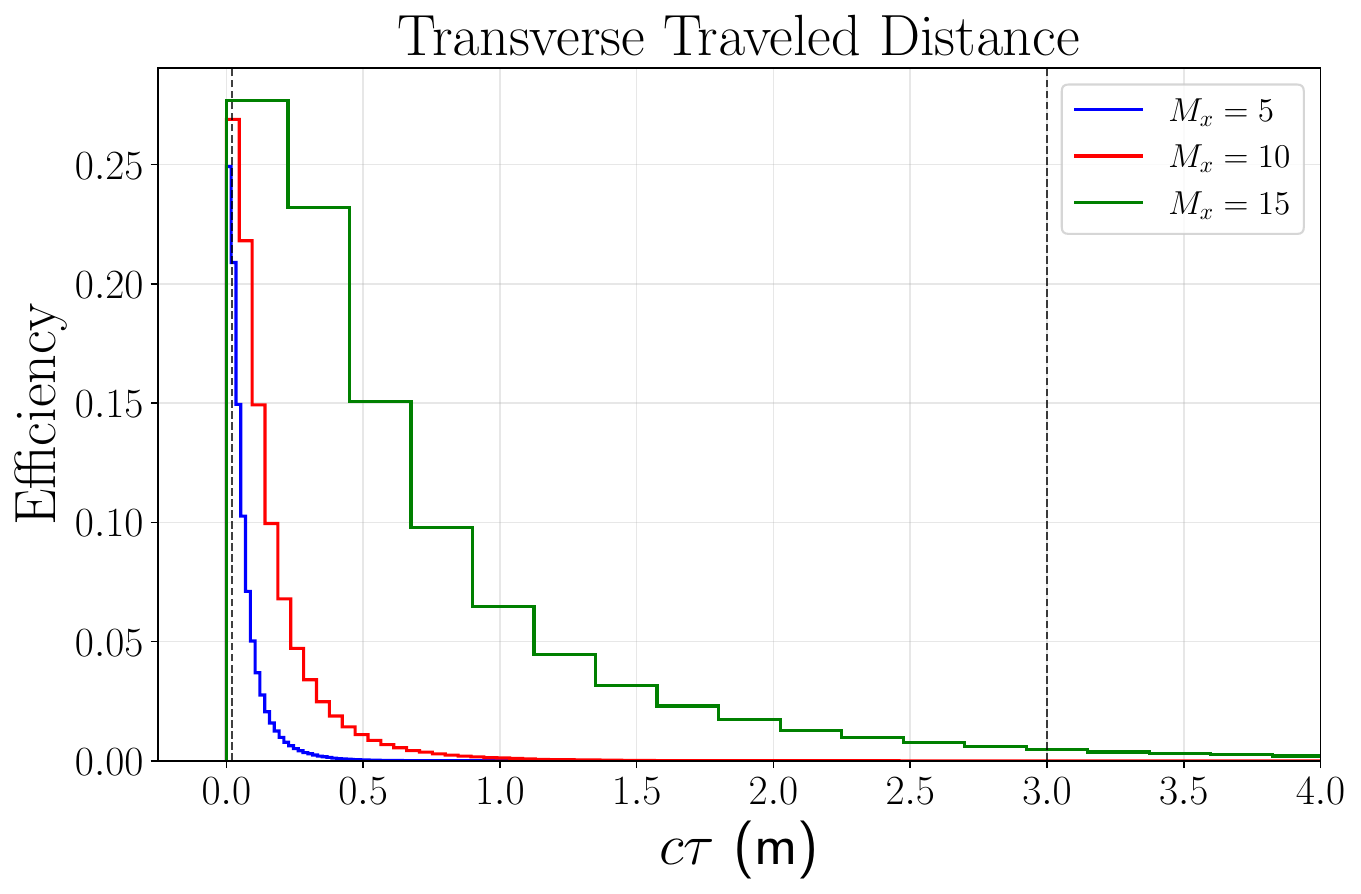}
    \end{subfigure}
    \caption{\small The selection efficiencies for signal processes as a function of lifetime. The dashed lines indicate our imposed limits for the X boson decay length.}
    \label{fig:traveled-distance}
\end{figure}

Figure~\ref{fig:traveled-distance} shows the signal selection efficiency as a function of the $X$ boson proper lifetime, 
with dashed lines indicating the geometric acceptance boundaries.

\subsection{Multivariate analysis}

To optimize signal–background separation, we employ a multivariate classifier based on adaptive
Boosted Decision Trees (BDTs) within the TMVA framework~\cite{Voss:2007jxm,Speckmayer:2010zz,Therhaag:2010zz}.
The multivariate strategy complements the baseline cut-based selection. While a cut-based analysis
carves the phase space via successive one- or two-sided thresholds, it can be suboptimal when
relevant observables exhibit nontrivial correlations. In contrast, decision trees recursively partition
the feature space into axis-aligned hyper-rectangles and assign each region to the signal or
background class. Boosting then aggregates many weak learners, yielding an effectively nonlinear
decision boundary in the multi-dimensional space that captures correlated structures and improves
discrimination power.

The BDT is trained using all simulated background categories and the signal samples, with events
weighted by their effective yields (cross section $\times$ integrated luminosity $\times$ generator
weight) so that the training reflects the expected composition after the baseline selection.
Independent training and test subsamples are used to monitor generalization. Overtraining is
checked using TMVA’s built-in tests: we compare the BDT score distributions for training and test
events for both classes and perform a Kolmogorov–Smirnov test; no evidence of overtraining is
observed within statistical uncertainties. 

The input variables are chosen to (i) capture the key physics differences between the LLP signal and
the Standard Model backgrounds (in particular heavy-flavor sources and quarkonium dimuons),
(ii) cover complementary aspects of the event topology (global kinematics, angular structure, and
track-based displacement), and (iii) avoid strong redundancy. Pairwise correlations were inspected
to ensure that the feature set is not dominated by a few highly correlated observables.

\subsubsection*{Discriminating Observables}
The set of observables fed into the BDT is designed to capture complementary aspects of the event topology:

\begin{itemize}
    \item MET-to-$H_T$ ratio ($\slashed{E}_T/H_T$): quantifies the balance between invisible and visible leptonic activity. 
    $H_T \equiv \sum_{\ell=e,\mu^\pm} p_T^\ell$ for the one electron and two muons in the event (jets excluded by design)
    \item Transverse mass of the $W$ boson ($m_T^W$): 
    $m_T^W = \sqrt{2\,p_T^e\,\slashed{E}_T\,\bigl(1 - \cos(\phi_e - \phi_{\slashed{E}_T})\bigr)}\,,$
    probing the kinematics of the prompt electron and the neutrino from the $W$ decay.
    \item Invariant mass of the muon pair ($M_{\mu^+,\mu^-}$): sensitive to quarkonium-like structures and the $X\to\mu^+\mu^-$ resonance shape.
    \item Invariant mass of the three-lepton system ($M_{e,\mu^+,\mu^-}$): encodes the overall leptonic topology of the signal.
    \item $\Delta R(e,\mu^-)$ and $\Delta R(e,\mu^+)$: separations in the $\eta$–$\phi$ plane, $\Delta R=\sqrt{(\Delta\eta)^2+(\Delta\phi)^2}$.
    \item $\cos\!\bigl(\Delta\phi(\mu^-,\mu^+)\bigr)$: azimuthal correlation between the two muons.
    \item $\cos\!\bigl(\Delta\phi(e,\mu^-)\bigr)$ and $\cos\!\bigl(\Delta\phi(e,\mu^+)\bigr)$: azimuthal correlations between the prompt electron and each muon.
    \item Transverse impact parameters $d_{xy}^{\mu^-}$ and $d_{xy}^{\mu^+}$: distances of closest approach of the muon tracks to the primary vertex in the transverse plane.
    \item Longitudinal impact parameters $d_{z}^{\mu^-}$ and $d_{z}^{\mu^+}$: distances of closest approach along the beam axis.
    \item Missing transverse energy ($\slashed{E}_T$): magnitude of the missing momentum in the transverse plane, primarily sensitive to neutrinos from $W\to\ell\nu$ decays.
    \item Scalar sum of lepton transverse momenta ($H_T$).
\end{itemize}

\begin{figure*}[!ht]
  \centering
    \begin{subfigure}[b]{0.32\textwidth} \includegraphics[width=\textwidth]{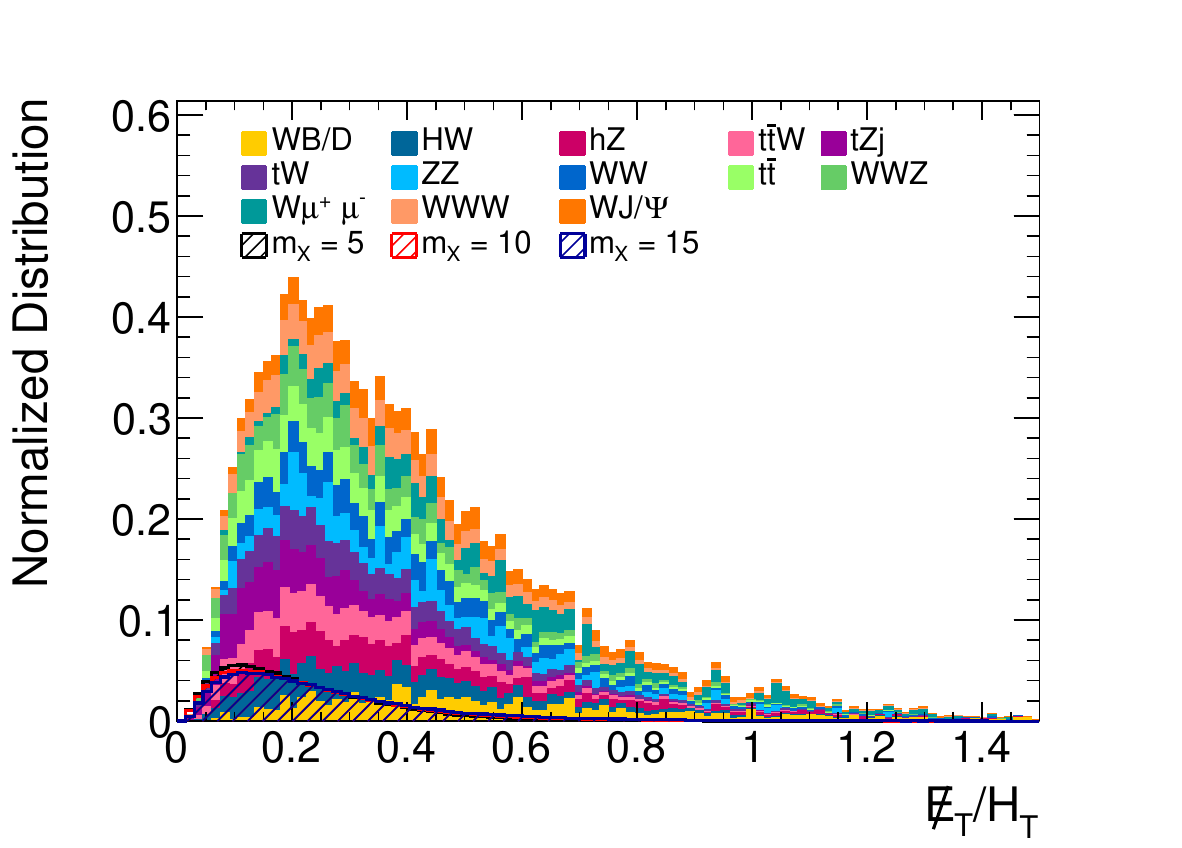} \caption{$\slashed{E}_T/H_T$} \end{subfigure}
    \begin{subfigure}[b]{0.32\textwidth} \includegraphics[width=\textwidth]{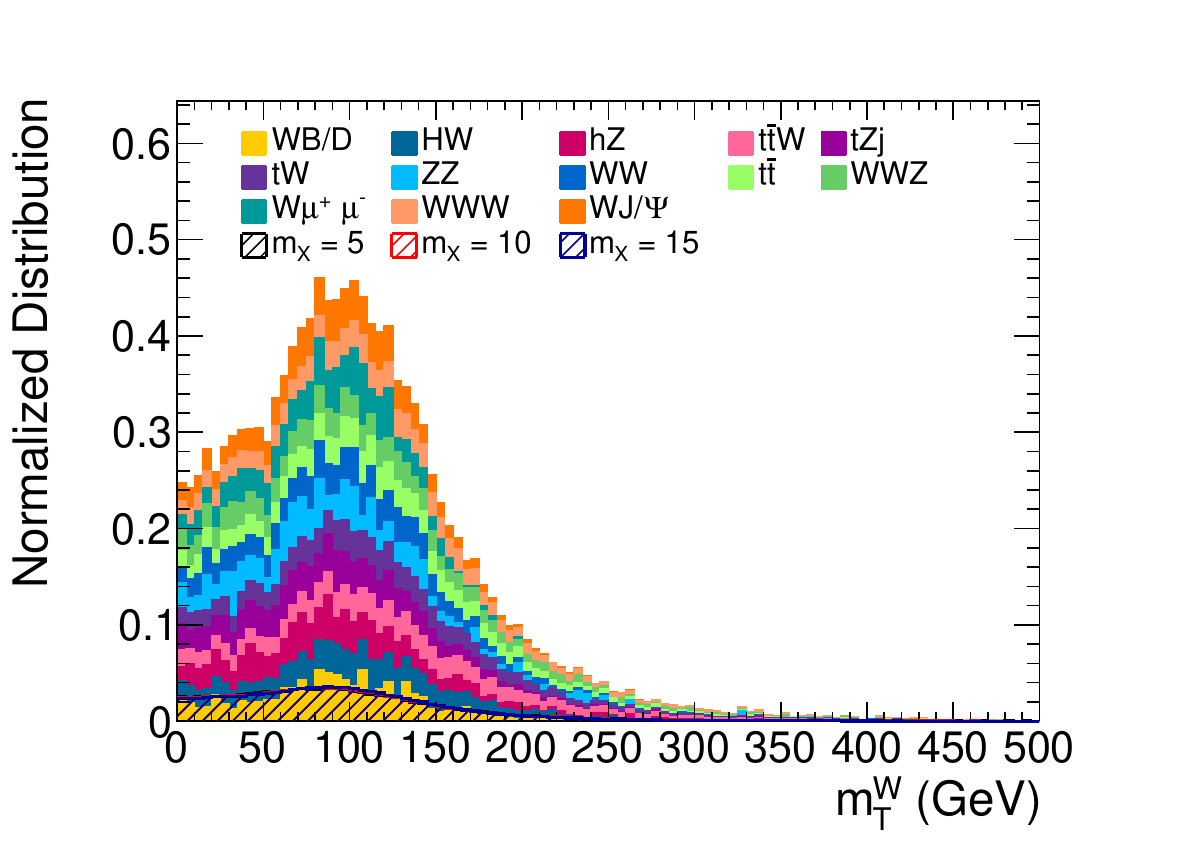} \caption{$m_T^W$} \end{subfigure}
    \begin{subfigure}[b]{0.32\textwidth} \includegraphics[width=\textwidth]{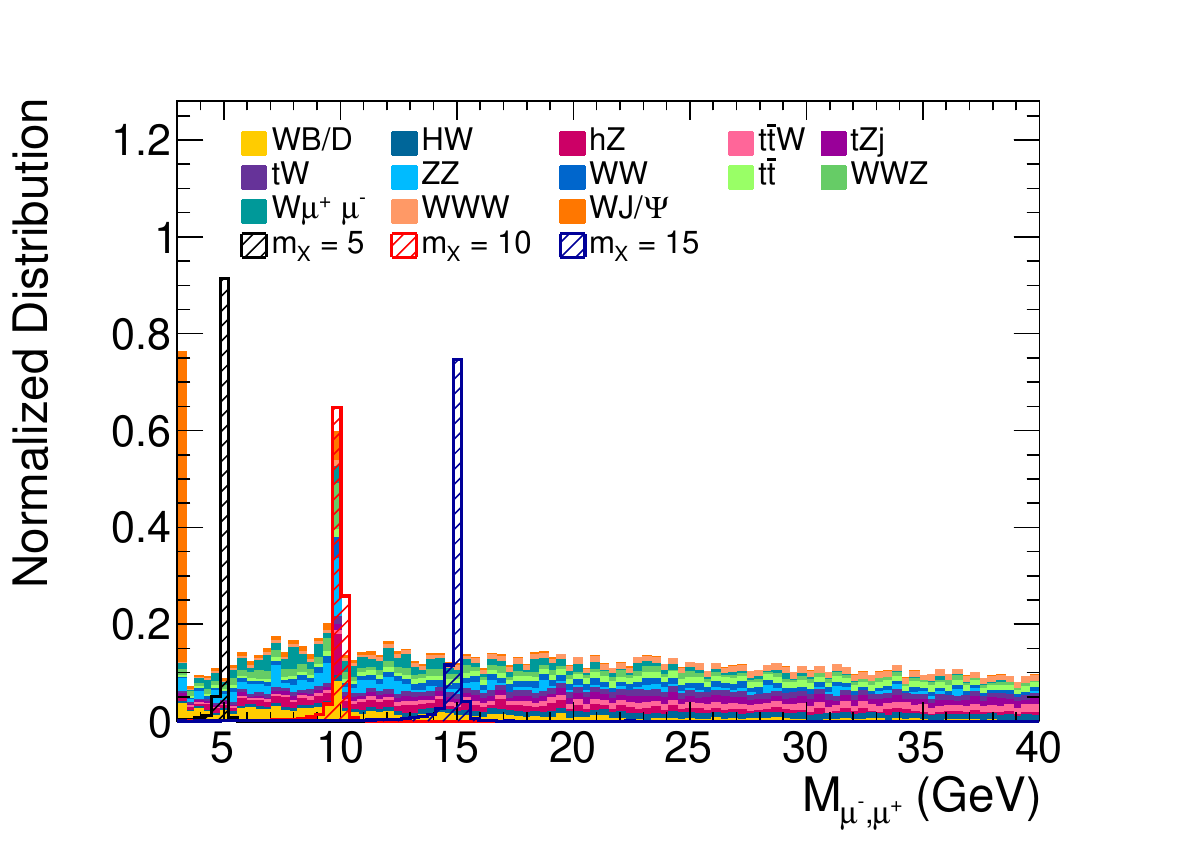} \caption{$M_{\mu^+,\mu^-}$} \end{subfigure}
    \begin{subfigure}[b]{0.32\textwidth} \includegraphics[width=\textwidth]{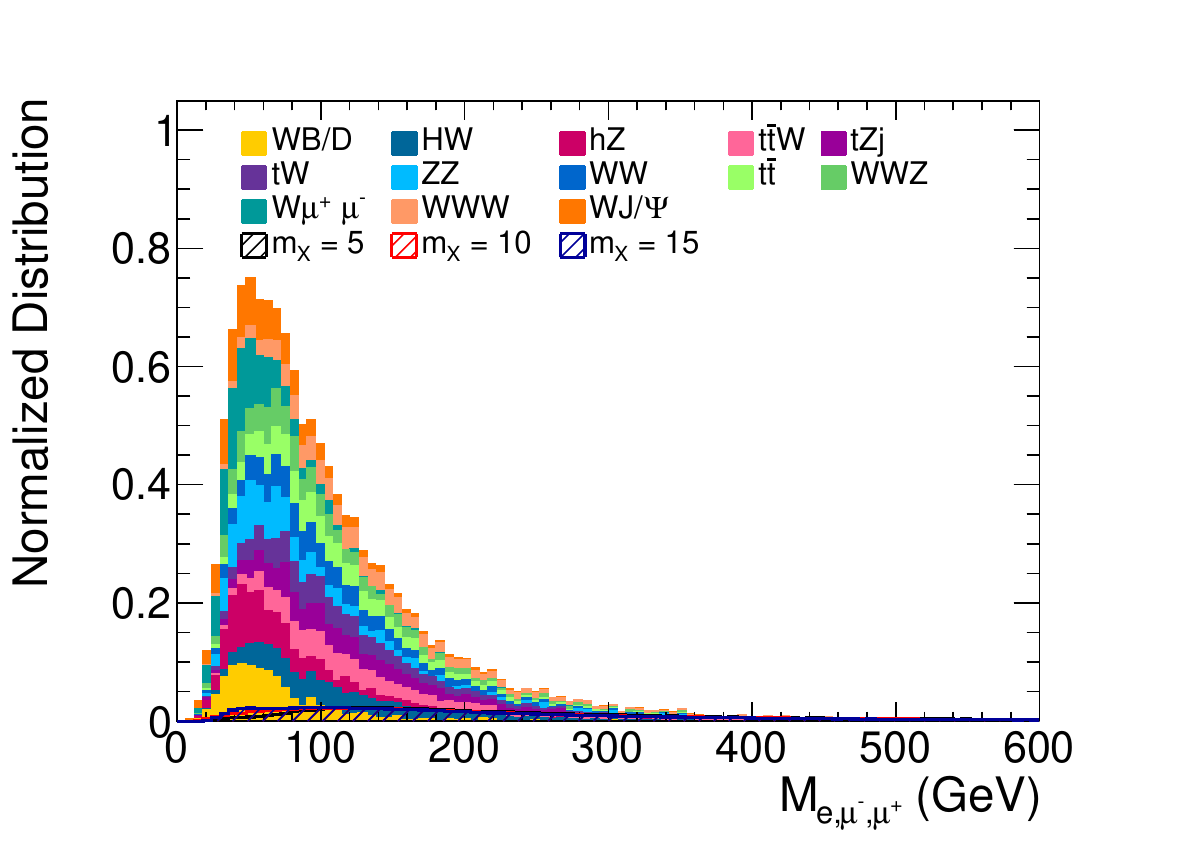} \caption{$M_{e,\mu^+,\mu^-}$} \end{subfigure}
    \begin{subfigure}[b]{0.32\textwidth} \includegraphics[width=\textwidth]{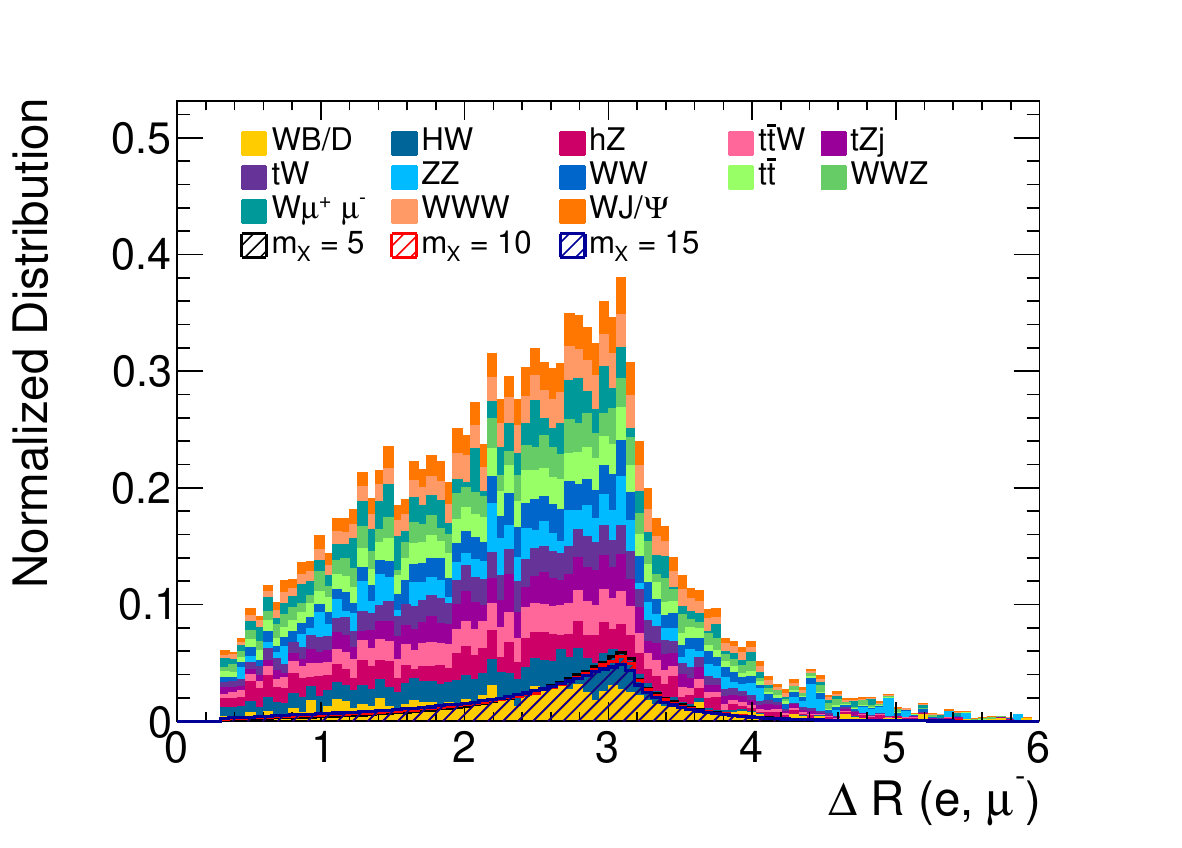} \caption{$\Delta R(e,\mu^-)$} \end{subfigure}
    \begin{subfigure}[b]{0.32\textwidth} \includegraphics[width=\textwidth]{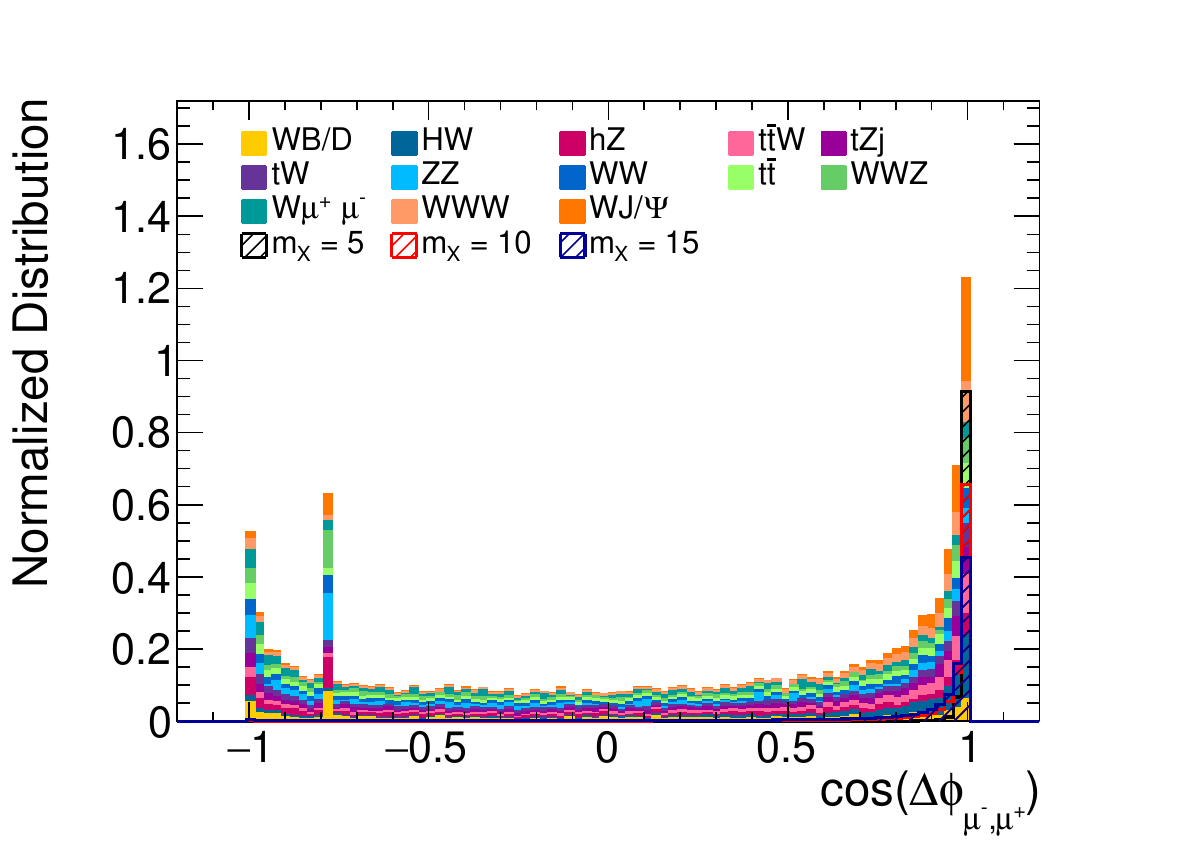} \caption{$\cos\Delta\phi(\mu^-,\mu^+)$} \end{subfigure}
    \begin{subfigure}[b]{0.32\textwidth} \includegraphics[width=\textwidth]{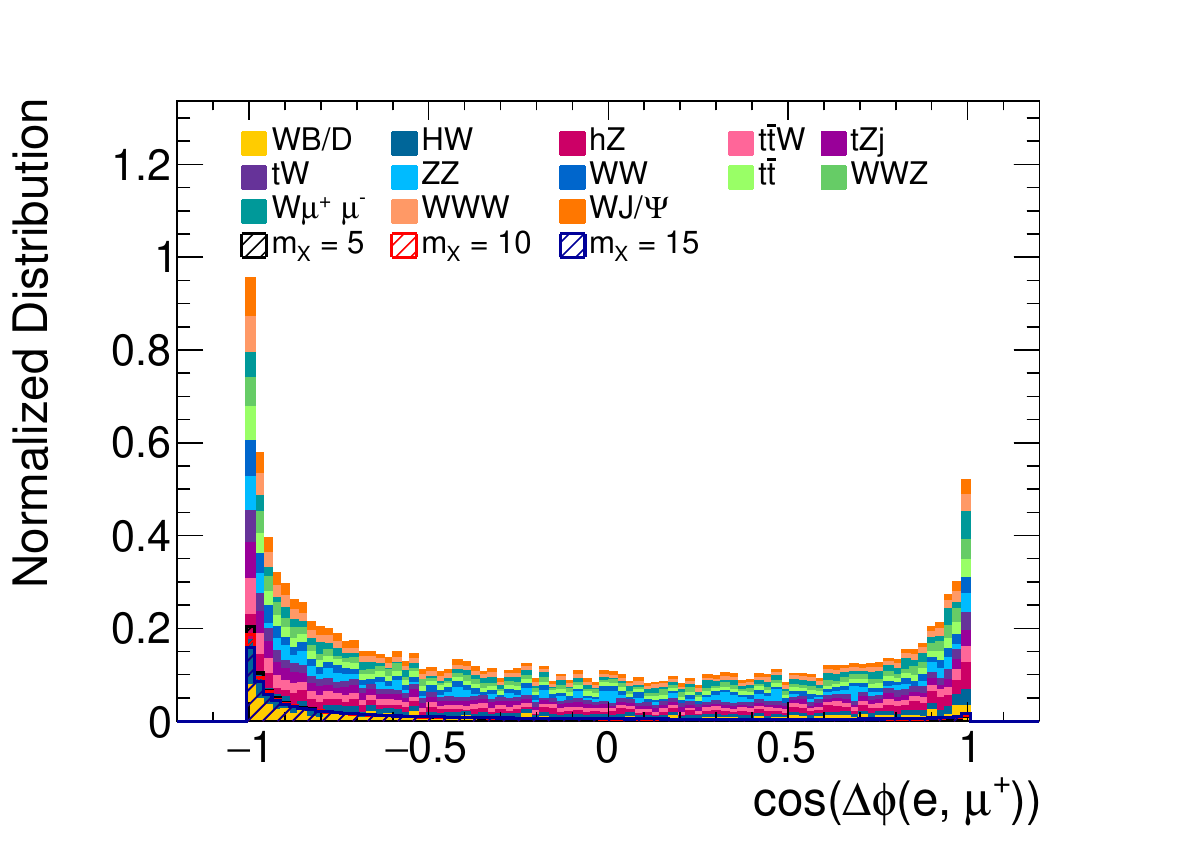} \caption{$\cos\Delta\phi(e,\mu^+)$} \end{subfigure}
    \begin{subfigure}[b]{0.32\textwidth} \includegraphics[width=\textwidth]{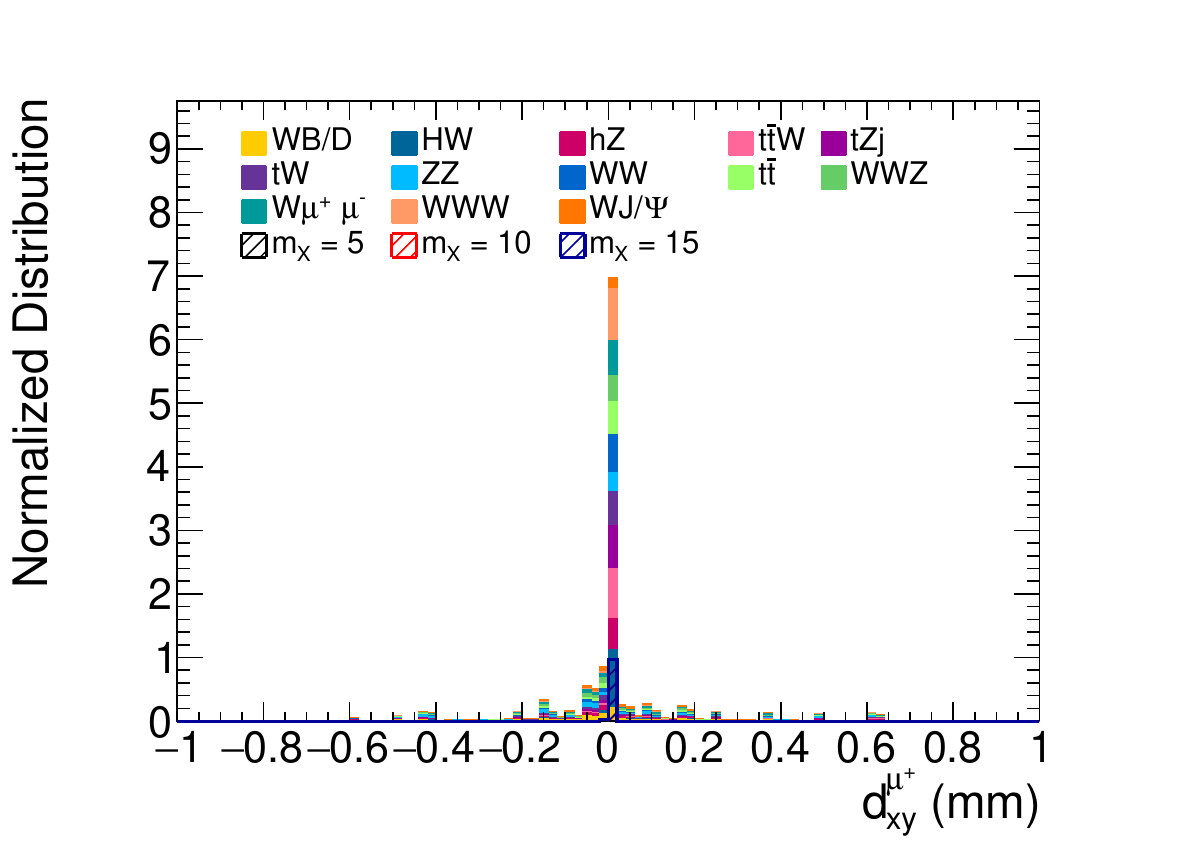} \caption{$d_{xy}^{\mu^+}$} \end{subfigure}
    \begin{subfigure}[b]{0.32\textwidth} \includegraphics[width=\textwidth]{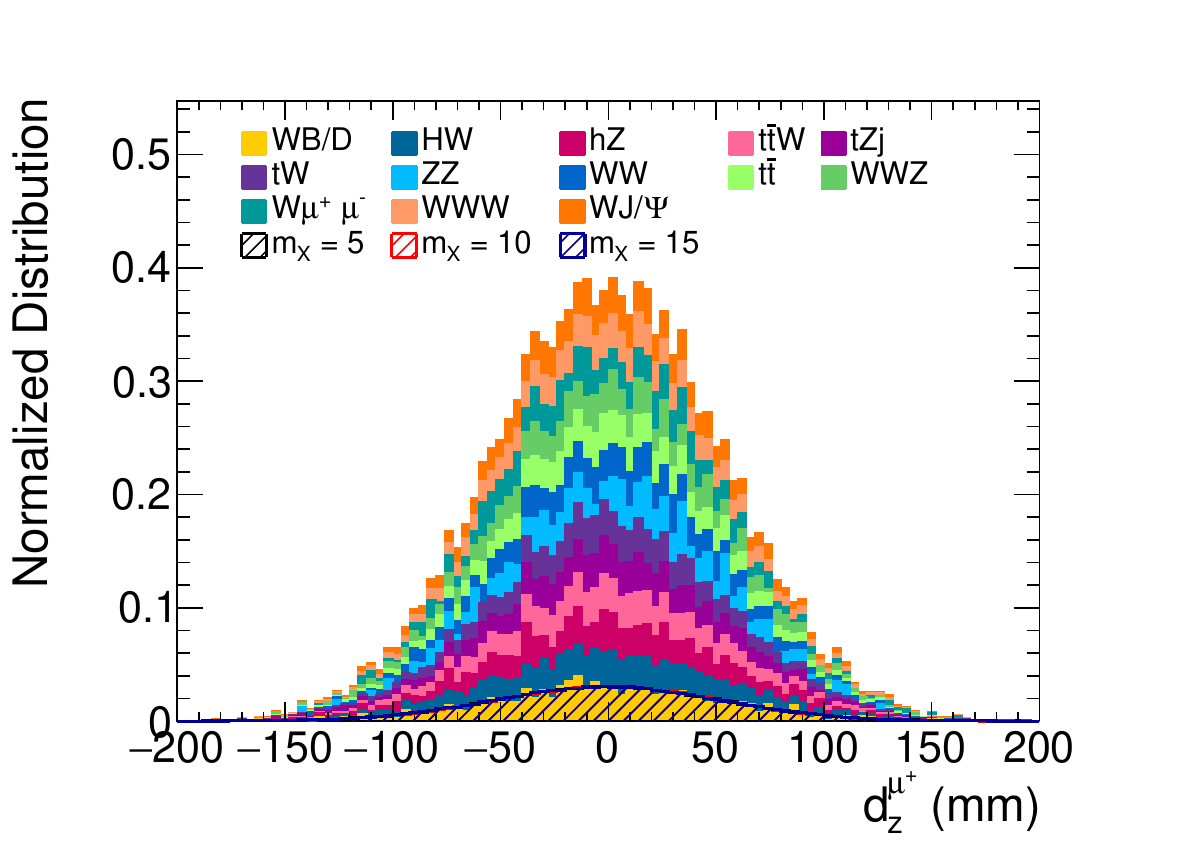} \caption{$d_{z}^{\mu^+}$} \end{subfigure}
    
  \caption{Normalized (unit-area) distributions of the input observables used in the BDT for the signal hypotheses $m_X=5,10,15~\mathrm{GeV}$ and 
  the dominant SM backgrounds, after the baseline selection. Variables span global kinematics, angular correlations, and displaced-track measurements.}
  \label{fig:distributions}
\end{figure*}

\begin{figure}[!ht]
\centering
    \begin{subfigure}[b]{0.48\textwidth} \includegraphics[width=\textwidth]{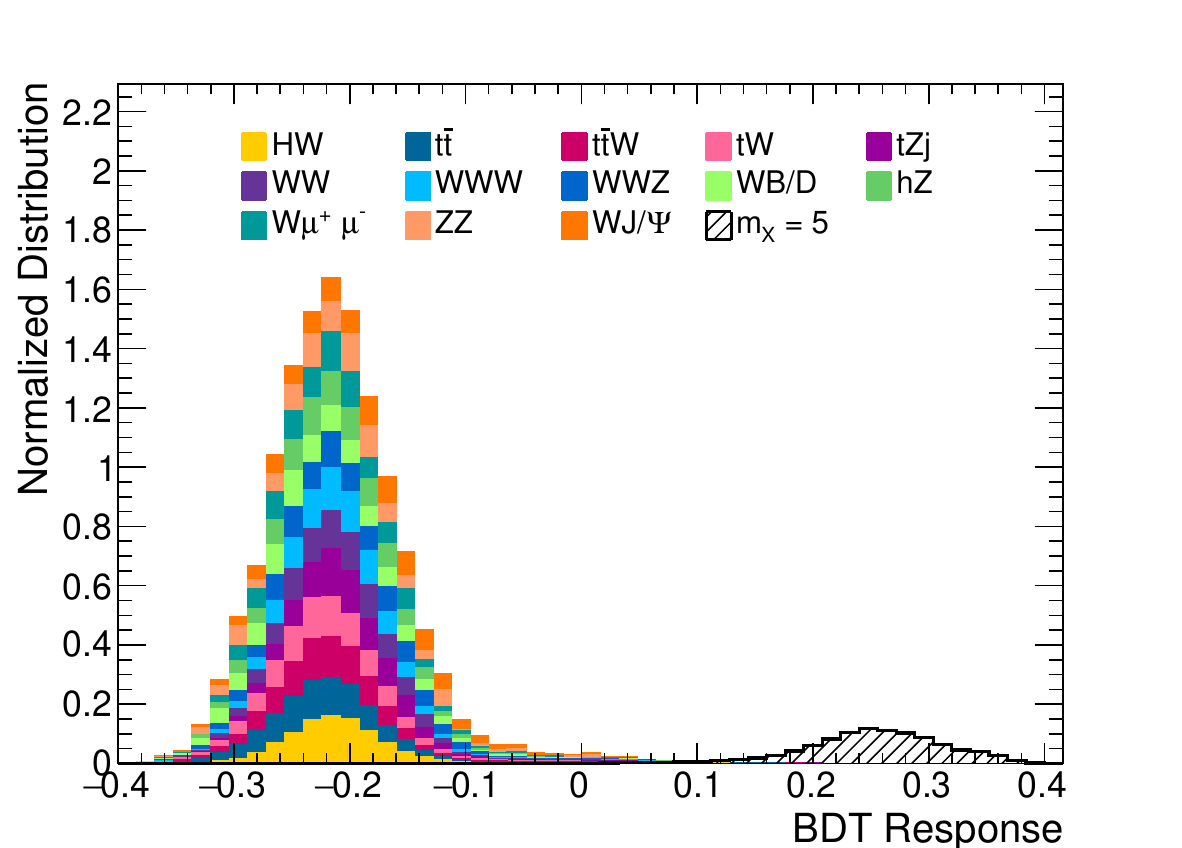} \caption{$m_X=5$~GeV} \end{subfigure}
    \begin{subfigure}[b]{0.48\textwidth} \includegraphics[width=\textwidth]{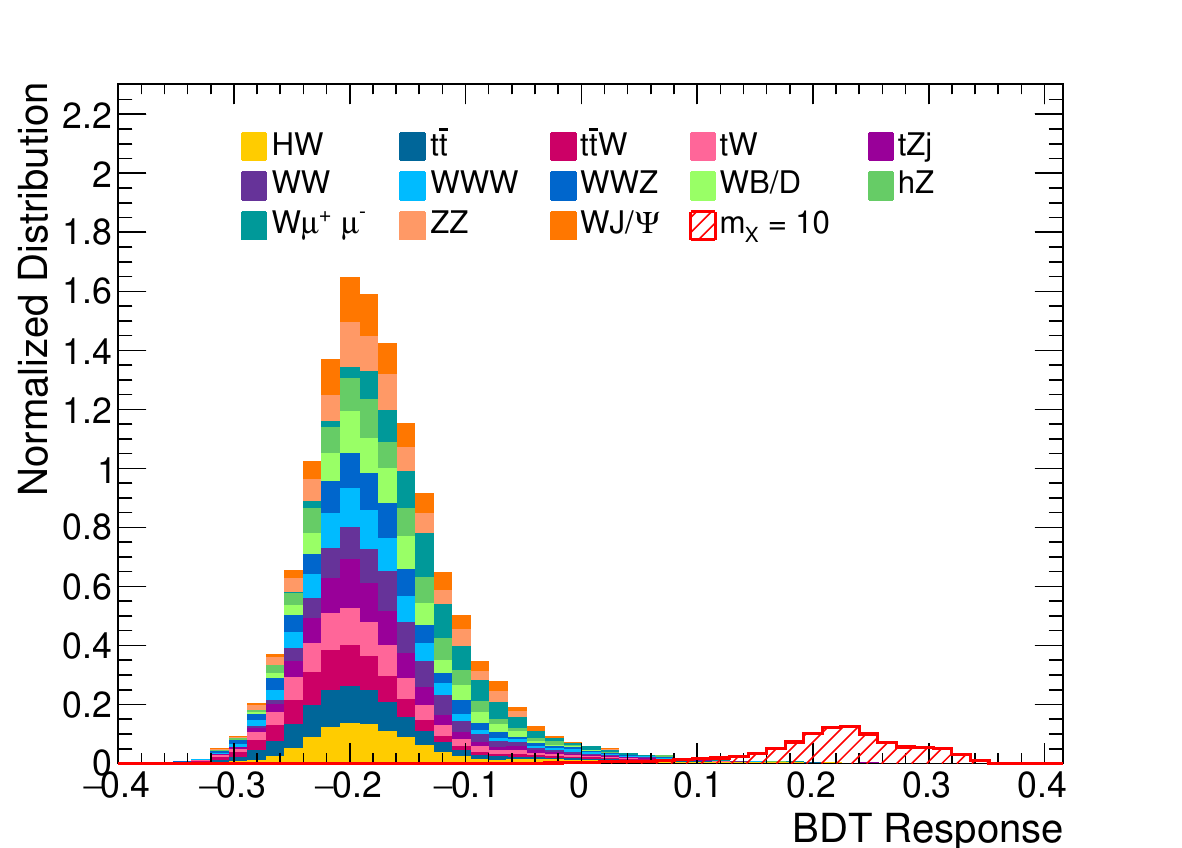} \caption{$m_X=10$~GeV} \end{subfigure}
    \begin{subfigure}[b]{0.48\textwidth} \includegraphics[width=\textwidth]{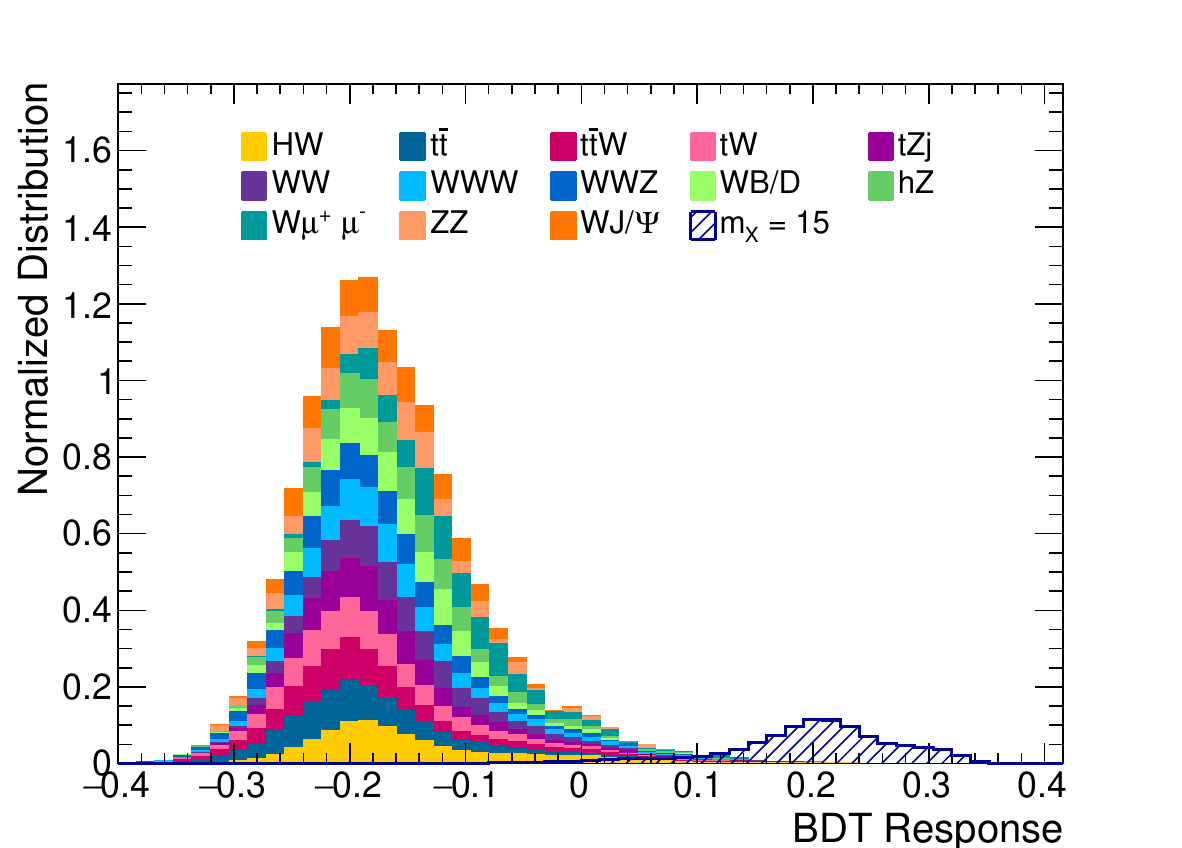} \caption{$m_X=15$~GeV} \end{subfigure}
    \caption{BDT score distributions for signal and background. Training and test samples agree within statistical uncertainties, indicating no overtraining.}
    \label{fig:bdt-response}
\end{figure}

The distributions obtained from the selected events for these observables are shown in 
Fig.~\ref{fig:distributions}, normalized to unit area for direct shape comparison between the signal
hypotheses ($m_X=5,10,15~\mathrm{GeV}$) and the dominant Standard Model backgrounds after the baseline selection.
The trained BDT responses for signal and background are shown in Fig.~\ref{fig:bdt-response}; 
the separation is stable across the three signal mass hypotheses.

\begin{figure}[!t]
  \centering
 \includegraphics[width=0.3\linewidth]{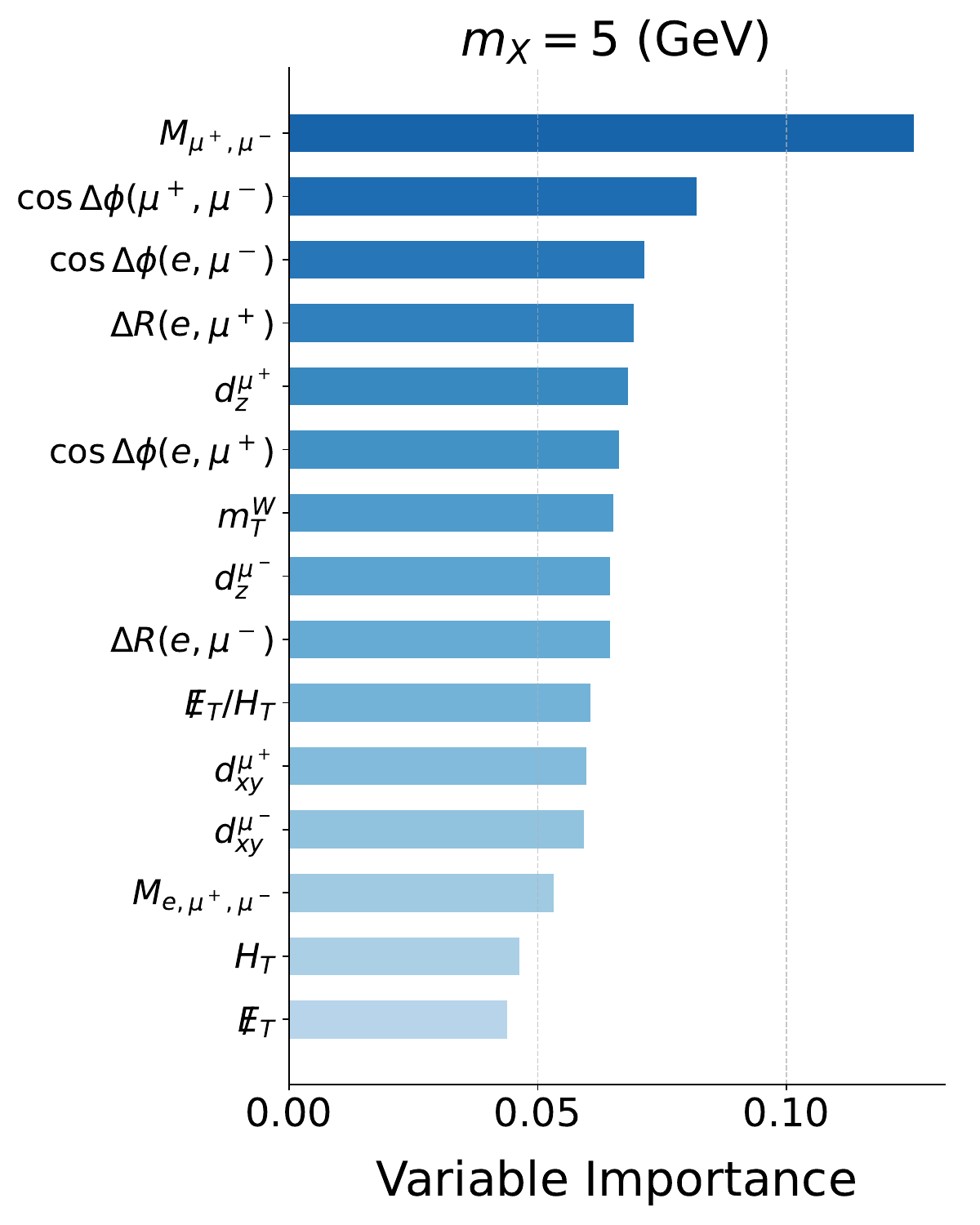}
  \includegraphics[width=0.3\linewidth]{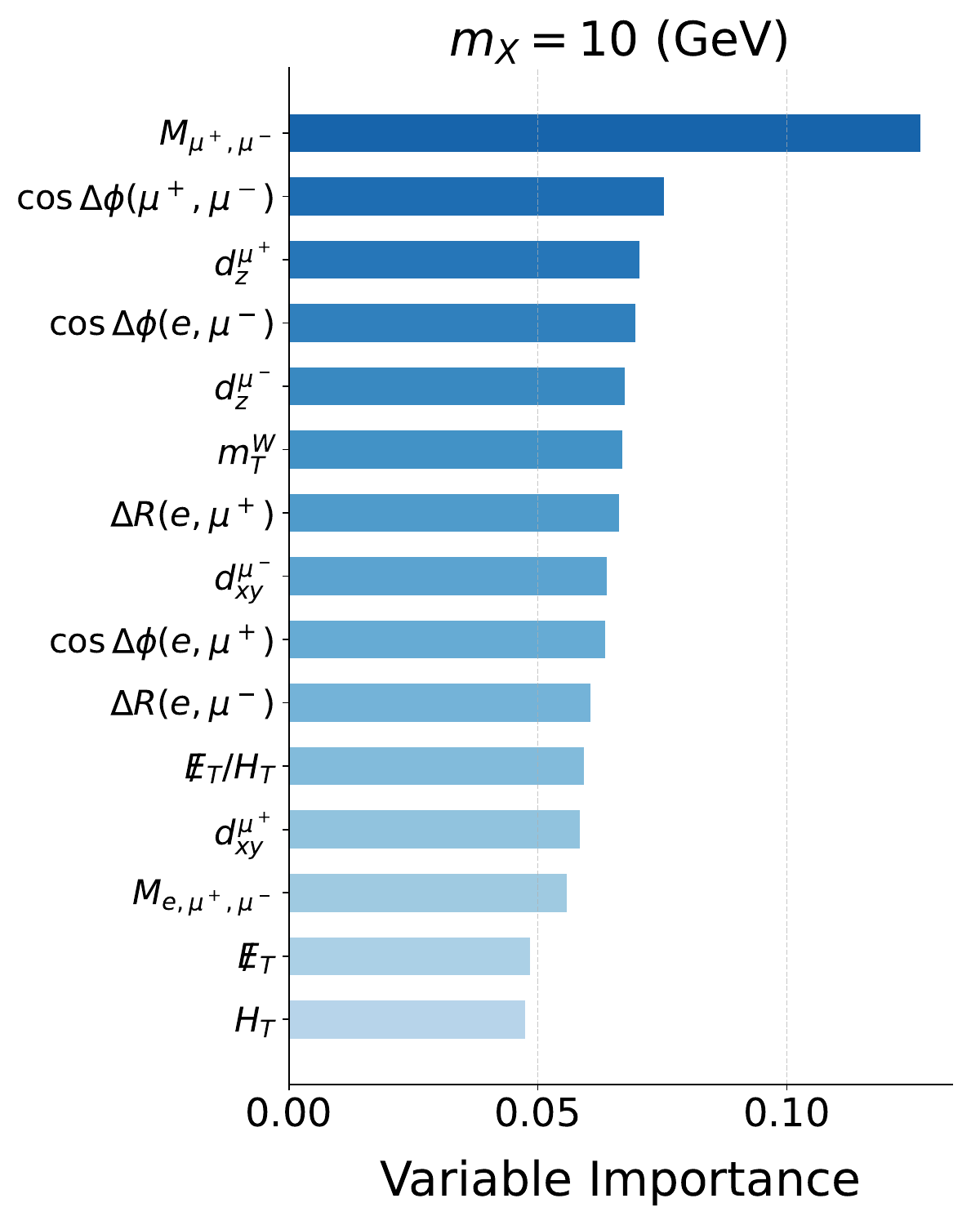}
 \includegraphics[width=0.3\linewidth]{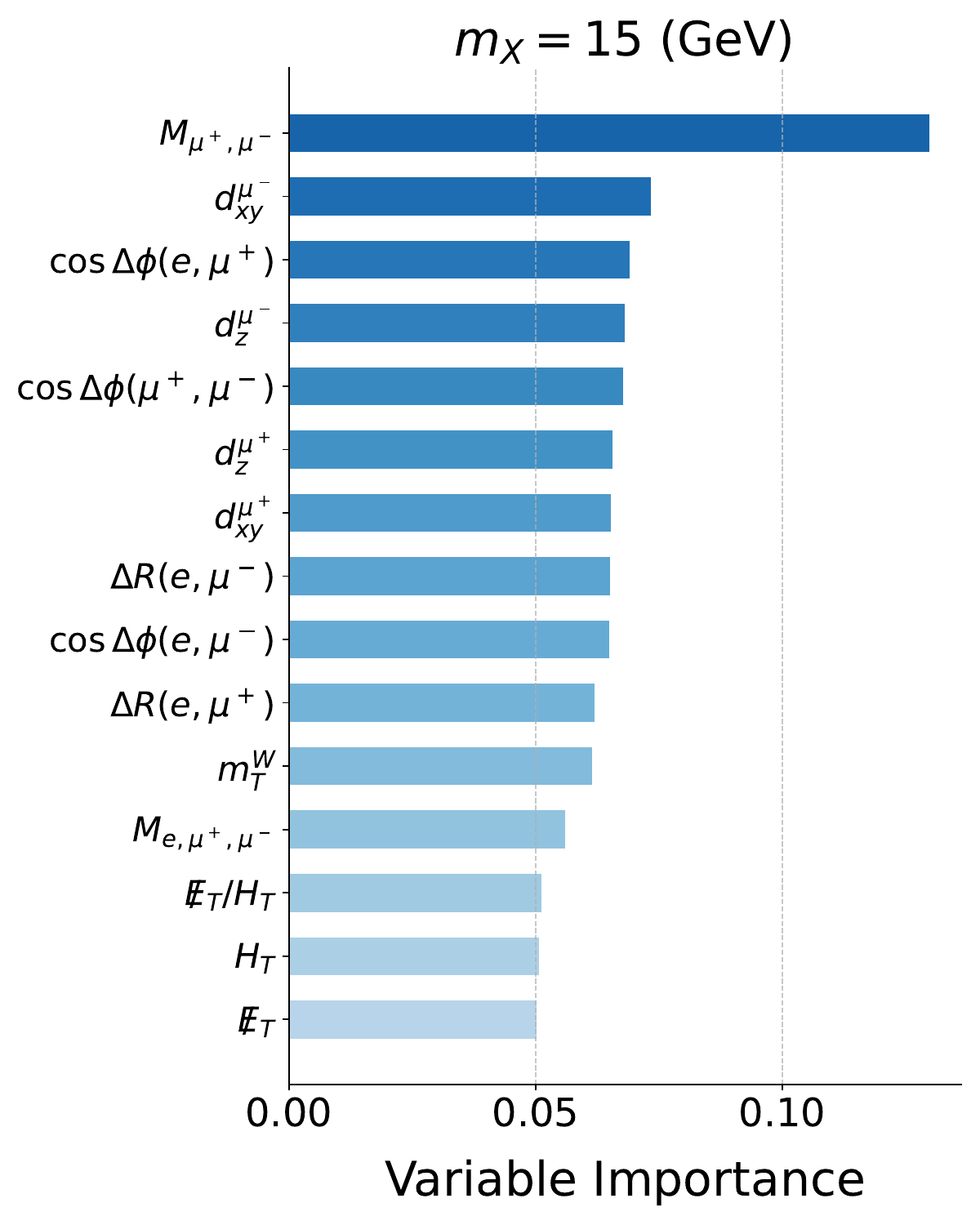}
  \caption{Relative importance of the input observables used in the BDT, evaluated with TMVA for the signal events with $m_X=5$ GeV, $m_X=10$ GeV, and $m_X=15$ GeV. 
  The ranking combines method-independent separation and the BDT gain metric. }
  \label{fig:bdt-ranking}
\end{figure}

The trained BDT score distributions for signal and background
demonstrate clear separation and consistent shapes between the training and test samples, indicating no overtraining.  
The relative importance of each observable, evaluated using TMVA’s method-independent separation and the BDT-specific gain metric, 
is summarized in Fig.~\ref{fig:bdt-ranking}.  
Both metrics identify displaced-track variables ($d_{xy}$, $d_z$), dimuon kinematics ($M_{\mu^+,\mu^-}$, $\cos\Delta\phi(\mu^-,\mu^+)$), and global event 
balance variables ( $m_T^W$, $\slashed{E}_T/H_T$) as the most powerful discriminants against heavy-flavor and quarkonium backgrounds.

\section{Results}\label{sec:results}

Using the response distributions from the BDT, we map out 95\% confidence level exclusion 
boundaries in the $g_{Xll}^2 - c_{W}^2$ parameter space, distinguishing regions consistent with background-only scenarios from those suggesting new physics signals. 
To set upper limits on the signal cross section and the corresponding model parameters, 
we employ a standard Bayesian approach~\cite{ParticleDataGroup:2024cfk, d0col}. 
A flat prior is assumed for the signal cross section. 
The likelihood function is constructed from the Poisson probability of observing $n$ events, 
given an expected number of background events $n_B$, a signal acceptance times efficiency 
$\varepsilon$, an integrated luminosity $\mathcal{L}_{\text{int}}$, and a signal cross section times branching fraction $S$:  
\begin{equation}
\mathcal{L}(n \, | \, S, \varepsilon, n_B, \mathcal{L}_{\text{int}}) \;=\; 
\frac{e^{-(n_B + \varepsilon S \mathcal{L}_{\text{int}})} \, (n_B + \varepsilon S \mathcal{L}_{\text{int}})^n}{n!} \, .
\end{equation}
Systematic uncertainties are incorporated through nuisance parameters, 
which are modeled with Gaussian priors reflecting their corresponding uncertainties. 
The posterior distribution for the parameter of interest, $S$, 
is obtained by marginalizing over the nuisance parameters. 
The 95\% confidence level upper limit on the signal cross section, $S^{95\%}$, 
is defined by solving  
\begin{equation}
\int_0^{S^{95\%}} \! \mathcal{L}(n \, | \, S) \, dS \;=\; 0.95 \, .
\end{equation}
This expected 95\% CL upper limit on $S$ is then translated into constraints on the parameters of the theoretical model.
To obtain more realistic projections, systematic uncertainties are incorporated into the analysis. The luminosity uncertainty is taken as $1.2\%$. 
Theoretical uncertainties on the signal arise from variations of the parton distribution functions, evaluated using 
the 100 NNPDF23~\cite{Ball:2012cx} error sets; from QCD scale variations, where the 
renormalization ($\mu_R$) and factorization ($\mu_F$) scales are independently changed by factors of 0.5, 1, and 2 using 
generator-level weights; and from modifications of the strong coupling constant $\alpha_S$ in the 
parton shower to account for initial- and final-state QCD radiation. 
In addition, an overall $5\%$ uncertainty on the background normalization and a $5\%$ uncertainty 
on the signal efficiency are included to account for potential experimental systematic effects not explicitly modeled in this study. 
A detailed assessment of these effects lies beyond the scope of the present work and should be addressed by the experimental collaborations.

\begin{figure}[!ht]
\centering
    \begin{subfigure}[b]{0.49\textwidth} 
        \centering
        \includegraphics[width=\textwidth]{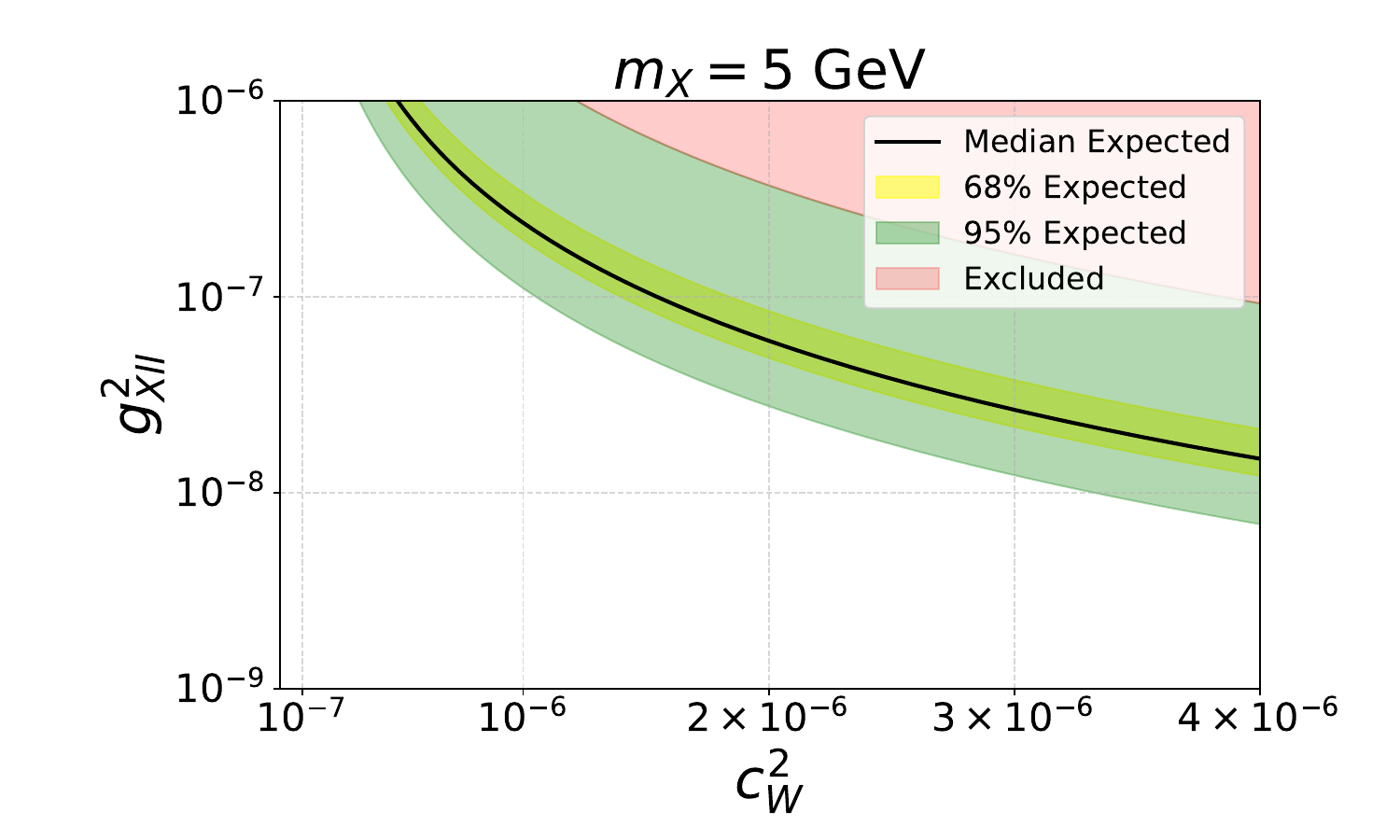}
        \caption{}
    \end{subfigure} 
    \begin{subfigure}[b]{0.49\textwidth} 
        \centering
        \includegraphics[width=\textwidth]{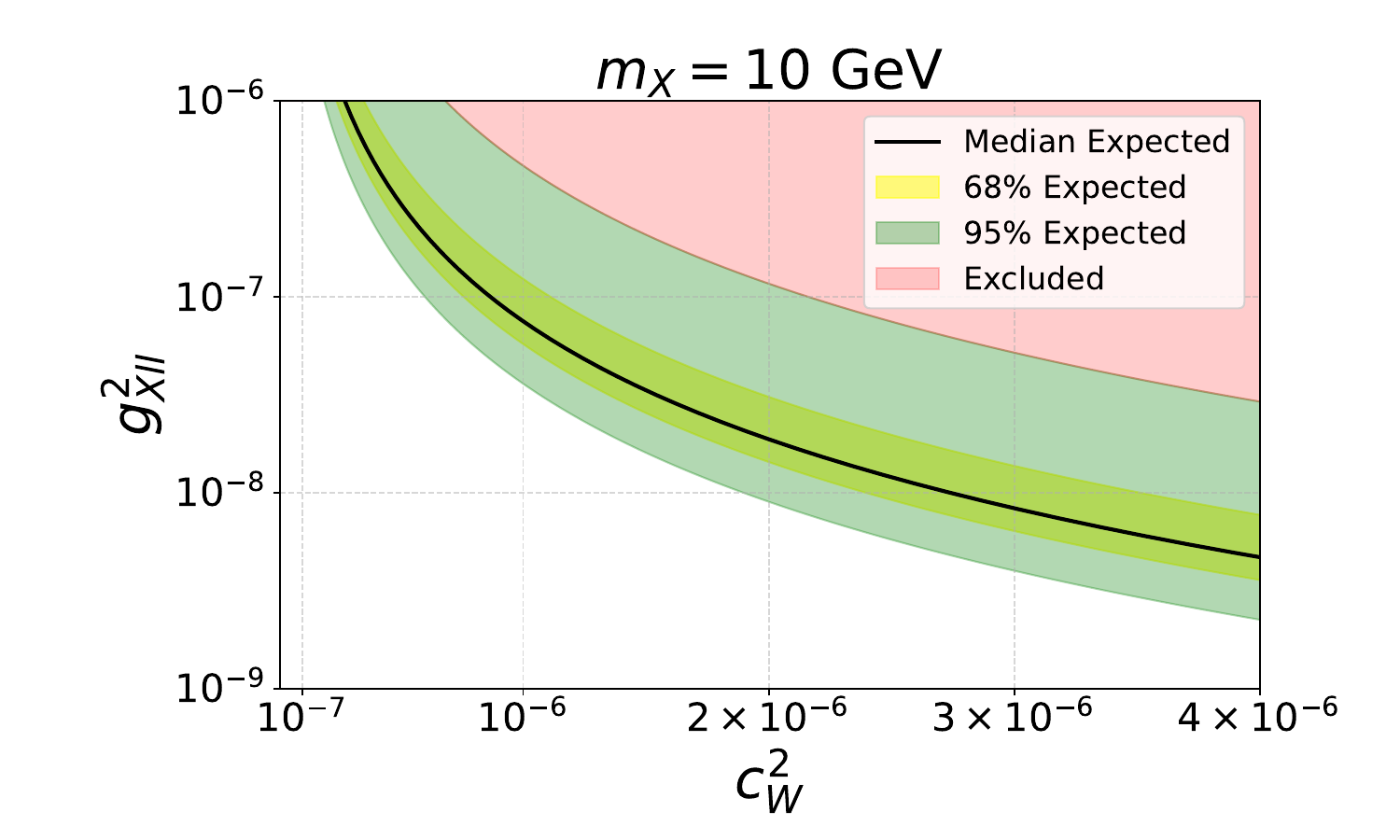}
        \caption{}
    \end{subfigure}
    \begin{subfigure}[b]{0.49\textwidth} 
        \centering
        \includegraphics[width=\textwidth]{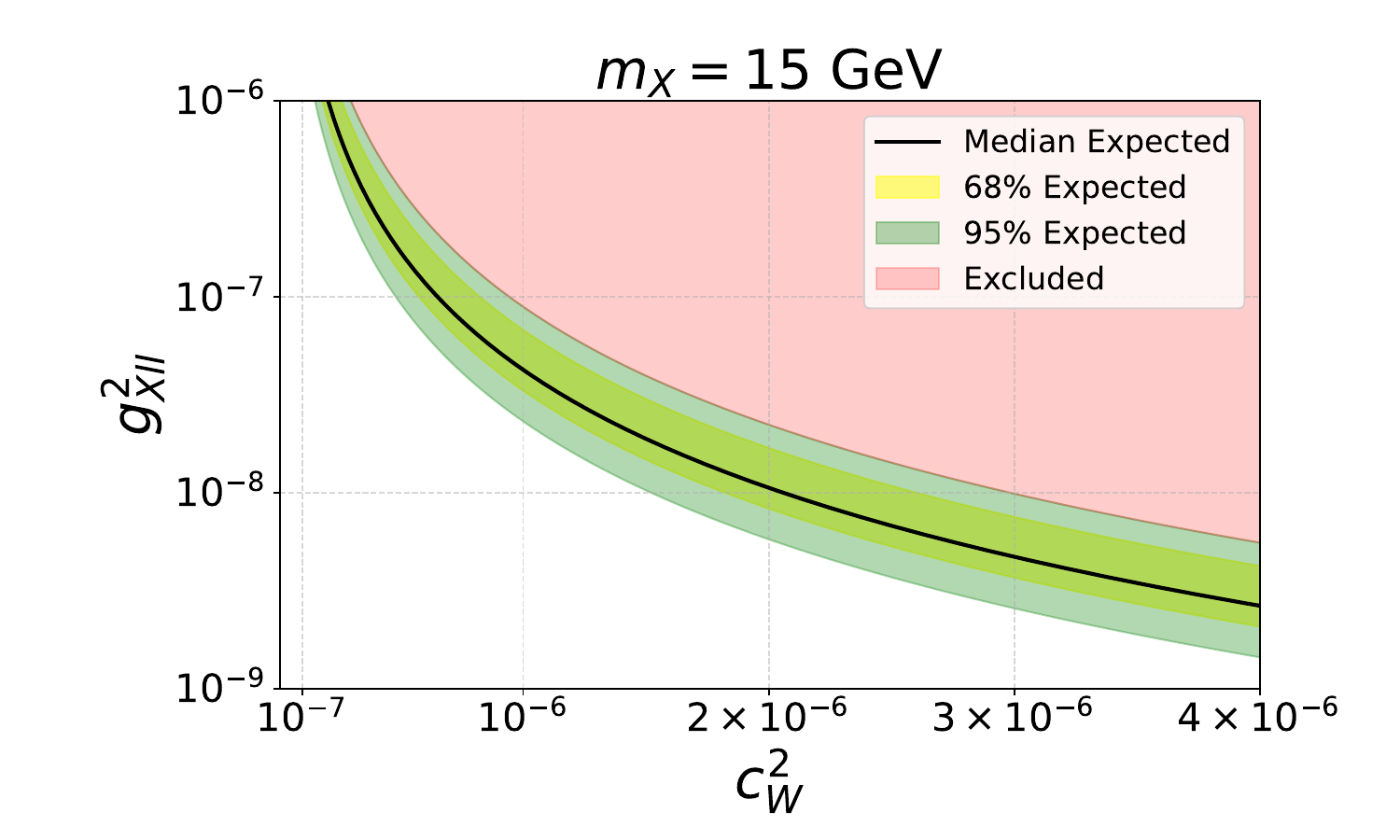}
        \caption{}
    \end{subfigure}
    \caption{\small Expected $95\%$ CL limits in the $g_{Xll}^2 - c_{W}^2$ plane for $m_X = 5$ GeV, $m_X = 10$ GeV, and $m_X = 15$ GeV signal processes. 
    The limits are based on the integrated luminosity of 3000 fb$^{-1}$ (HL-LHC) and include both statistical and systematic uncertainties.}
    \label{fig:exclusion}
\end{figure}

Figure~\ref{fig:exclusion} presents the expected $95\%$ CL exclusion contours in the $g_{Xll}^2 - c_{W}^2$ 
parameter space for representative LLP mass hypotheses of $m_X = 5$, $10$, and $15~\text{GeV}$. 
The limits are derived for an integrated luminosity of 3000~fb$^{-1}$ at the HL-LHC, incorporating all relevant 
statistical and systematic effects. The results demonstrate that, even under conservative systematic assumptions, 
the analysis achieves strong sensitivity to previously unexplored regions of parameter space, 
substantially extending the reach beyond current constraints. 
These projections set a robust benchmark for future LLP searches in multi-lepton final states with displaced dimuons.

In Ref.~\cite{Bondarenko:2019tss}, a long displaced-vertex  reconstruction scheme was proposed, 
demonstrating sensitivity to $X$ boson mass of $m_X = 10~\mathrm{GeV}$ and couplings down to $c_W^2 \sim 10^{-8}$ for $g_{X\ell\ell}^2 \sim 10^{-6}$.  
The parameter space probed in that study lies well below the experimental bound on $c_W$, namely $c_W^2 \lesssim 10^{-3} \, (m_X / 1~\mathrm{GeV})^2$~\cite{Alekhin:2015byh}.
In our analysis, for $g_{X\ell\ell}^2 \sim 10^{-7}$, we exclude values of $c_W^2$ above $10^{-6}$ 
for $m_X = 5~\mathrm{GeV}$ and above $\sim 10^{-7}$ for $m_X = 15~\mathrm{GeV}$ at 95\% CL.  
As shown in Fig.~\ref{fig:exclusion}, the sensitivity improves with increasing $m_X$ in the range studied, with the strongest constraints obtained for $m_X = 15~\mathrm{GeV}$.  
This trend reflects the enhanced kinematic separation between signal and background at higher $X$ boson masses, leading to more effective multivariate discrimination.
The results presented in Ref.~\cite{Bondarenko:2019tss} are obtained in a setup optimized to explore the ultimate reach at the HL-LHC, 
whereas our work incorporates some systematic uncertainties, the pileup conditions expected at the HL-LHC, realistic detector effects, 
and, as far as possible, the relevant background processes. 
This leads to more conservative but experimentally robust sensitivities, 
offering a complementary perspective to the projections of Ref.~\cite{Bondarenko:2019tss}.


\section{Summary and Conclusions}\label{sec:conclusion}

We have investigated the discovery and exclusion potential of the HL-LHC at $\sqrt{s} = 14$~TeV for 
the Chern–Simons portal, an extension of the SM featuring a massive neutral vector boson $X$ 
associated with an additional $U_X(1)$ gauge symmetry. In this framework, gauge anomaly cancellation is 
realized via topological Chern–Simons  terms in the low-energy effective action, generated by integrating out 
heavy chiral fermions charged under both the SM and $U_X(1)$. The resulting mixed anomalies (e.g.\ $U_X(1)SU(2)^2$) 
lead to unsuppressed CS couplings at the TeV scale, giving rise to distinctive collider signatures.

Our study focuses on the associated production of the $X$ boson with a $W$ boson and jets, yielding a 
final state with two oppositely charged muons, one electron, missing transverse energy, and at least one jet. The analysis employs a Boosted Decision Tree 
to optimize signal–background separation, incorporating realistic detector effects and systematic uncertainties. 
These include luminosity, parton distribution functions, QCD scale variations, and QCD radiation modeling, 
as well as additional conservative uncertainties of 5\% on the background normalization and 5\% on 
the signal efficiency to account for residual experimental effects.
Assuming an integrated luminosity of 3000~fb$^{-1}$, we derive expected 95\% CL exclusion contours in the $g_{X\ell\ell}^2 - c_W^2$ parameter 
space for benchmark $X$ boson masses of 5, 10, and 15~GeV. 
The projected sensitivity extends significantly beyond existing limits, demonstrating that the HL-LHC can 
probe previously unexplored regions of parameter space, particularly for higher $X$ boson masses. 
The results highlight the HL-LHC’s unique potential to explore the Chern–Simons portal, 
providing complementary  constraints to other experimental approaches, 
and setting the stage for future searches in multi-lepton final states with displaced dimuons.


\end{document}